\documentclass[reqno]{amsart} \usepackage{amscd}
\usepackage{epsfig}
\newtheorem{theorem}{Theorem}[section]
\newtheorem{proposition}[theorem]{Proposition}
\newtheorem{lemma}[theorem]{Lemma}
\newtheorem{corollary}[theorem]{Corollary}
\theoremstyle{definition}
\newtheorem{definition}[theorem]{Definition}

\theoremstyle{remark} \newtheorem{remark}[theorem]{Remark}
\numberwithin{equation}{section}
\newcommand{\be}{\begin{eqnarray}} 
\newcommand{\ee}{\end{eqnarray}}
\newcommand{\field}[1]{\ensuremath{\mathbb{#1}}}
\newcommand{\CC}{\field{C}}

\newcommand{\RR}{\field{R}}

\DeclareMathOperator{\D}{D(SU(2))}
\DeclareMathOperator{\T}{{\mathcal T}}
\DeclareMathOperator{\Tr}{Tr}

\DeclareMathOperator{\SU}{SU}

\begin{document}

\title[Particles and GFT]{Quantum Gravity with Matter \\ via Group Field Theory}
\author{Kirill Krasnov} \address{School of Mathematical Sciences \\
  University of Nottingham \\ Nottingham, NG7 2RD, UK}

\begin{abstract} A generalization of the matrix model idea to quantum
gravity in three and higher dimensions is known as group field theory
(GFT). In this paper we study generalized GFT models that can be
used to describe 3D quantum gravity coupled to point particles. The
generalization considered is that of replacing the group leading to pure
quantum gravity by the twisted product of the group with its dual
--the  so-called Drinfeld double of the group. The Drinfeld double is
a quantum group in that it is an algebra that is both non-commutative
and non-cocommutative, and special care is needed to define group
field theory for it. We show how this is done, and study the
resulting GFT models. Of special interest is a new topological model
that is the ``Ponzano-Regge'' model for the Drinfeld double. However, as we show, 
this model does not describe point particles. Motivated by the 
GFT considerations, we consider a more general class of models that 
are defined using not GFT, but the so-called chain mail techniques. 
A general model of this class does not produce 3-manifold invariants, 
but has an interpretation in terms of point particle Feynman diagrams.  
\end{abstract}
\maketitle
\section{Introduction}
\label{sec:intr}

Recently a question of coupling of quantum gravity to matter has been revisited in works
\cite{Barrett,Freidel}. Both papers deal with quantum gravity in 2+1 dimensions, and analyze how
matter Feynman diagrams are modified when the quantum gravity effects are
taken into account. Both work contain results that are intriguing. However, certain
conceptual issues were left unclear. Thus, work \cite{Barrett} has suggested
that the effect of quantum gravity is in modifying the measure of integration 
$\prod_v dx_v$ in a Feynman amplitude given by:
\be\label{coord-rep}
{\mathcal Z} = \int \prod_v dx_v \prod_{\langle vv'\rangle} G(x_v-x_{v'}).
\ee
Here one integrates over the positions of vertices $x_v$ of a Feynman graph, $G(x-y)$ is 
the particle propagator in coordinate representation, and $\langle vv'\rangle$ denotes 
edges of the Feynman diagram. In this scheme quantum gravity is only responsible for
a modification of the $\prod_v dx_v$ measure, and the particle content as well as the
Lagrangian must be specified independently. In particular, one is not forced to using any particular
propagator. Thus, this proposal is that of ``quantum gravity + matter''.
Work \cite{Freidel}, on the other hand, showed that particular quantum
gravity amplitudes can be interpreted as particle Feynman diagrams, provided
one makes a special choice of the particle propagator. Thus, in this
interpretation ``quantum gravity = matter''. These two interpretations
seem to be in conflict. This paper is devoted to an analysis of these and 
related issues. We will argue that both interpretations are correct.

The basic philosophy that is to be pursued in the present paper is as follows. 
Consider the momentum representation Feynman amplitude of some
theory. The precise definition of the theory is unimportant for us at
this stage. We shall restrict our attention to general aspects common
to Feynman amplitudes of all theories. Thus, the amplitude is given by:
\be\label{mom-rep}
{\mathcal Z} = \int \prod_{\langle vv'\rangle} dp_{\langle vv'\rangle} G(p_{\langle vv'\rangle})
\prod_v \delta(\sum_{v'} p_{\langle vv'\rangle}).
\ee
Here the integrals are taken over the momenta $p_{\langle vv'\rangle}$ on each 
edge of the digram; in addition, one has a momentum conservation law for each vertex.
We have to modify the Feynman amplitude \eqref{mom-rep} so that
it describes point particles. Point particles in 2+1 dimensions are 
conical singularities in the metric. As we shall see, this has the effect that
particle's momentum is {\it group-valued}. For point particles in 
Euclidean 3-dimensional space, which is what we consider in
this work, the relevant group is $\SU(2)$. The proposal to be developed in this paper is
that point particles and processes involving them can still be described by Feynman diagrams \eqref{mom-rep},
but the momentum $p$ on each edge of the diagram should be taken group-valued. The rest of
the paper is devoted to various tests of this proposal. 

Some point particle theories we consider are topological; we refer the
reader to the main body of the paper for the precise definition of
what this means. As we shall see, these topological particle theories
will be produced by the GFT. Other theories we consider are
non-topological; GFT is of no direct relevance here. However, the
techniques developed in the GFT context (such as the chain mail, see
below) will prove extremely useful for these theories as well. This
is how the GFT is omnipresent throughout the paper.

We start by reminding the reader how point particles in 3D are described. For simplicity,
this paper deals only with the case of $3+0$-dimensional spacetime, that is, with 3D metrics
of Euclidean signature, and the cosmological constant is set to zero.

\subsection{Point particles}

In 3 spacetime dimensions presence of matter has much more drastic consequences than in 4D: 
the asymptotic structure of spacetime gets modified. This has to do with the fact that the vacuum Einstein
equations are much stronger in 3D: they require that the curvature is constant (zero) 
everywhere. Then an asymptotically flat spacetime is flat everywhere. When matter is 
placed inside it introduces a conical defect that can be felt at infinity. The modification
due to matter is in the leading order, not in the next to leading as in higher dimensions.
This has the consequence that the total mass in spacetime is proportional to the deficit angle
created at infinity. Because the angle deficit cannot increase $2\pi$, the mass is
bounded from above. All this is very unlike the case of higher dimensions, and makes
the theory of matter look rather strange and unfamiliar. But such strange features are
at the end of the day responsible for a complete solubility of the theory.

Point particles give the simplest and most natural form of matter to consider in 3D. A point
particle creates a conical singularity at the point where it is placed. The angle deficit
$2\theta$ at the tip of the cone is related to particle's mass as $4\pi M G=\theta$. Point
particles move along geodesics. Their worldliness are lines of conical singularities.

To make this description more explicit, let us consider how a space containing a line
of conical singularity can be obtained. This is achieved by performing
a rotation of
space, and identifying points related by this rotation. A rotation is described by an
element $g(2\theta)\in\SU(2)$, which we choose to parameterize as follows:
\be
g(\theta)=e^{i \theta \vec{n}\cdot\vec{\sigma}}= \cos{\theta} + i \vec{n}\cdot\vec{\sigma} \sin{\theta}.
\ee
Here $\vec{n}: \vec{n}\cdot\vec{n}=1$ is a unit vector. It describes the direction of
the axis of rotation, and thus the direction of particle's motion. The angle $2\theta$
specifies the amount of rotation, and thus gives the mass of the particle. The reason for
choosing $\theta$ and not $2\theta$ for parameterizing the particle's mass will become clear below. 
An equivalent description in more mathematical terms is to say that a particle of a given mass $\theta$
is described by a group element $g\in C_\theta$, where $C_\theta$ is a conjugacy class in $\SU(2)$.
Non-trivial conjugacy classes $C_\theta, \theta\not=0$ in $\SU(2)$ are spheres $S^2$
and a point in $C_\theta$ describes the direction of motion. Thus, the information about
both the mass and the direction of motion is encoded in $g$. This is why this quantity plays
the role of momentum of a point particle. Note that the group-valued nature
of particle's momentum automatically incorporates the bound on its mass. Note also
that not only the upper, but the lower bound is naturally imposed as well --
the $\SU(2)$ valued momenta are those of non-negative mass particles.

Once points of $\RR^3$ related by a rotation $g$ are identified, we obtain a space with a 
line of conical singularity in the direction $\vec{n}$. The angle deficit is $2\theta$.
The space has zero curvature everywhere except along particle's worldline. One can compute the
holonomy of the spin connection around the worldline and verify that it is equal to $g(\theta)$. It is for
this reason that the quantity $\theta$ equal to half the angle deficit gives a more convenient measure
parameterization of the mass. Let us now consider two point particles, with group elements $g_1,g_2$ 
describing them. The holonomy around the pair is the product of holonomies, and the mass 
parameter $\theta$ of the pair is given by:
\be\label{mass-comb}
\cos{\theta} = \frac{1}{2} \Tr(g_1 g_2) = \cos{\theta_1}\cos{\theta_2} - \vec{n_1}\cdot\vec{n_2} 
\sin{\theta_1} \sin{\theta_2}.
\ee
Note that it does matter now in which order the product of holonomies is taken, for 
$g_1 g_2\not= g_2 g_1$ as we are dealing with elements of the group. But in determining the mass
of the combined system this ambiguity is irrelevant, for the trace is taken. Let us 
look at \eqref{mass-comb} in more detail. In case $\vec{n_1}$ is parallel to $\vec{n_2}$, this formula
reduces to: $\theta=\theta_1+\theta_2$. When particles move in opposite directions we have:
$\theta=|\theta_1-\theta_2|$. Thus, the angle $\theta$ is more properly interpreted as
the magnitude of particle's momentum, not as its kinetic energy. 

It is instructive to compare \eqref{mass-comb} to the usual law of addition of vectors in $\RR^3$.
Thus, consider two particles with momenta $\vec{p_1}, \vec{p_2}$. Define: $m=|\vec{p}|, \vec{n}=\vec{p}/|\vec{p}|$.
Then we have:
\be\label{flat-comb}
m^2 = m_1^2 + m_2^2 + 2 \vec{n_1}\cdot\vec{n_2} m_1 m_2.
\ee
This formula for addition of momenta is to be compared with \eqref{mass-comb}. Indeed, as $\theta=4\pi MG$,
\eqref{mass-comb} can be expanded in powers of $G$. This gives exactly \eqref{flat-comb}, modified by
$O(G^2)$ corrections:
\be
M^2 = M_1^2 + M_2^2 + 2 \vec{n_1}\cdot\vec{n_2} M_1 M_2 + \\ \nonumber
\frac{4\pi^2 G^2}{3}\left( M^4 - M_1^4 - M_2^4 - 3M_1^2 M_2^2 - 4 \vec{n_1}\cdot\vec{n_2} M_1 M_2 (M_1^2+M_2^2)
\right) +O(G^4).
\ee
Thus, the fact that the law \eqref{mass-comb} of addition of momenta has the correct limit as $G\to 0$
gives additional support to our identification of $g$ with particle's momentum.

Having understood why particle's momentum becomes group valued, let us remind the reader 
some basic facts about the Ponzano-Regge model
\cite{PR} of 3D quantum gravity, and how point particles can be 
coupled to it.

\subsection{Ponzano-Regge model in presence of conical singularities}

In Ponzano-Regge model,
one triangulates the 3-manifold in question (``spacetime''), and assigns 
a certain amplitude to each triangulation. This amplitude turns out to be triangulation
independent and gives a topological  invariant of the 3-manifold. Let us
remind the reader how the Ponzano-Regge amplitude \eqref{PR} can be
obtained from the gravity action. When written in the first order formalism, zero cosmological constant
3D gravity becomes the so-called BF theory. The action is given by:
\be\label{bf}
S[{\bf e},{\bf a}]=\int_M {\rm Tr}({\bf e}\wedge {\bf f}).
\ee
Here $\bf e$ is a Lie algebra valued one form, and $\bf f$ is the curvature of a $G$-connection $\bf a$
on $M$. The path integral of this simple theory reduces to an integral over flat connections:
\be\label{Z-PR}
{\mathcal Z}[M]=\int {\mathcal D}{\bf e} {\mathcal D}{\bf a}\,\, e^{iS[{\bf e},{\bf a}]}=\int {\mathcal D}{\bf a}\,\,
\delta({\bf f}).
\ee
A discretized version of this last integral can be obtained as follows. Let us triangulate the
manifold $M$ in question. Let us associate with every edge of the dual triangulation the 
holonomy matrix for the connection $\bf a$. This holonomy matrix is an element of the group $G$. The
product of these holonomies around the dual face must give the trivial element (because the connection is flat).
Thus, one can take the $\delta$-function of a product of group elements around each dual face,
multiply these $\delta$-functions and then integrate over the group elements on dual edges.
This is the discretized version of \eqref{Z-PR}. By decomposing the $\delta$-function on the
group $G=\SU(2)$ into characters, it is easy to show that this
procedure gives the following amplitude:
\be\label{PR-n}
\sum_{\{j_e\}} \prod_e {\rm dim}(j_e) \prod_t (6j).
\ee
It turns out, however, that this amplitude is not yet triangulation
independent. In order to make it such one has to introduce a certain
formal prefactor. Thus, let us consider the following sum over
irreducible representations of $\SU(2)$:
\be\label{eta}
\eta^{-2}:= \sum_j {\rm dim}_j^2.
\ee
Note that this sum diverges and the quantity $\eta$ defined
by it is formal. The sum above is equal to the $\delta$-function on the
group evaluated at the identity element. To make the amplitude
\eqref{PR-n} triangulation independent one has to consider the
following multiple of it:
\be
\label{PR}
\eta^{2V} \sum_{\{j_e\}} \prod_e {\rm dim}(j_e) \prod_t (6j),
\ee
where $V$ is the number of vertices in the triangulation used to
obtain \eqref{PR-n}. The amplitude \eqref{PR} is now (formally,
because of a prefactor that is actually equal to zero) triangulation
independent. 

The origin of this formal prefactor is now well-understood, see
\cite{Freidel-div}. It has to do with the fact that the action
\eqref{bf} in addition to the usual $\SU(2)$ gauge symmetry has a 
non-compact gauge-symmetry. Indeed, because of
the Bianchi identity $d_A F=0$ the action is invariant under the
following transformation: $B\to B+ d_A \phi$. Integration over $B$
in the path integral includes integration over these gauge directions
and produces infinities that has to be cancelled by the zero prefactor 
in \eqref{PR}. A more systematic procedure for dealing with these
divergences was developed in \cite{Freidel-div,Freidel-CS} and
requires a certain gauge fixing procedure. We shall not consider such
a gauge fixing here and will proceed at a formal level, just keeping
track of the (formal) pre-factors $\eta$. All amplitudes can be
rendered well-defined by passing to the quantum group context, that
is, by considering the quantum group $\SU_q(2)$ instead of
$\SU(2)$. When $q$ is a root of unity the cut-off on the spin present
in the representation theory of $\SU_q(2)$ makes the sum that defines
$\eta^{-2}$ finite, and all the invariants well-defined. One should
always have this passage to $\SU_q(2)$ in mind when considering the
ill-defined Ponzano-Regge model expressions.

We have explained how the amplitude in \eqref{PR} can be obtained from an integral over flat connections.
It is now easy to add a point particle. Consider the case when a point particle is present and (a segment of)
its worldline
coincides with one of the edges of the triangulation of $M$. It will then create a conical singularity
along this edge. The product of holonomies around the face dual to this edge is not
trivial anymore. Instead it will lie in a conjugacy class determined by
the mass of the particle (by its deficit angle). Thus, for dual edges that encircle particles'
worldline one should put a $\delta$-function concentrated on the conjugacy class determined by
the particle. This gives a modified amplitude, with presence of the point particle taken
into account. Let us introduce a $\delta$-function that is picked on a particular conjugacy
class $\theta$:
\be
\delta_\theta(x) := \int dg \delta_e(g h_\theta g^{-1} x) = \sum_j \chi_j(h_\theta) \chi_j(x).
\ee
Here $\delta_e(x)$ is the usual $\delta$-function on the group picked
at the identity element. The sum in the second formula is taken over
all irreducible  representations of the group $\SU(2)$, and
$\chi_j(g)$ are the characters. The special group element $h_\theta$ is a representative
of the conjugacy class $\theta: h_\theta={\rm diag}(\exp{i\theta},\exp{-i\theta})$. Note that when $\theta=0$,
the above $\delta$-function reduces to the usual $\delta$-function picked at the identity. Let us now
assume that a point particle is present along all the edges of the triangulation, and
denote the angle deficit on the edge $e$ by $\theta_e$. The modified quantum 
gravity amplitude in the presence of particles is given by:
\be\label{PR-pp}
\sum_{\{j_e\}} \prod_e \chi_{j_e}(h_{\theta_e}) \prod_t (6j).
\ee
If particles are present along all the edges, there is no need to
introduce a formal $\eta$ prefactor as the amplitude is not required
to be triangulation independent. If the particle is only present
along some of the edges, one gets what can be called an observable of
Ponzano-Regge model, see \cite{Barrett-Obs}. In this case pre-factors
of $\eta$ for vertices away from the particle edges are necessary to
ensure a partial triangulation independence, see \cite{Barrett-Obs}
for more details. 

As it was explained in \cite{Barrett} and \cite{Freidel}, the
amplitude \eqref{PR-pp}
is the response of the quantized gravitational field to particle's presence. In the setting
described the particle is present as en external source in the gravitational path 
integral. The natural question is if one can integrate over the
particle degrees of freedom and thus have a quantum theory of particles coupled to
quantum gravity. In other words, the question is how to ``second quantize'' the
particle degrees of freedom. Some sketches on how this can be done were presented in both 
\cite{Barrett} and \cite{Freidel}. What is missing in these papers is a principle
that would generate Feynman diagrams for both gravity and particle degrees of
freedom. This paper is aimed at filling this gap. As we shall see, the
Feynman amplitudes for some of the theories (namely the topological
ones) can be generated by GFT. To understand
how this is possible, let us remind the reader some basic facts about the usual field
theory Feynman diagrams, and, in particular, about a certain duality transformation that can be
performed on a diagram. Let us remark that for the most part this paper deals with vacuum 
(closed) Feynman diagrams only. A generalization
to open diagrams that are essential for doing the scattering theory will be 
left for future work.

\subsection{Momentum representation and duality}

In addition to the coordinate \eqref{coord-rep} and momentum \eqref{mom-rep} representations
of Feynman diagrams there is another, the so-called dual representation. Let us
remind the reader how the dual representation is obtained in the case of
two spacetime dimensions. To get the dual formulation we need to endow 
Feynman graphs with an additional ``fat'' structure. Namely, each edge
of the digram must be replaced by two lines. The fat structure is equivalent
to an ordering of edges at each vertex. The lines of the fat graph are then
connected at vertices as specified by ordering. This structure is also
equivalent to specifying a set of 2-cells, or faces of the diagram. Once this
additional structure is introduced, one can define the {\it genus} $G$ of the diagram
to be given by the Euler formula: $2-2G=V-E+F$, where $V,E,F$ are the numbers
of vertices, edges and faces of the diagram correspondingly. Note that the same
Feynman diagram can be given different fat structures and thus have different
genus. For example, the diagram with 2 tri-valent vertices and 3 edges (the $\theta$-graph)
can correspond to both genus zero and genus one after the fat structure is
specified. 

Having introduced the fat structure, we can solve all the momentum
conservation constraints. It is simplest to do it in the case of $G=0$, so let us 
specialize to this situation. Let us introduce the new momentum variables,
one for every face of the diagram. Let us denote these by $p_f$. Given a pair
$\langle ff'\rangle$ of adjacent faces, there is a unique edge $\langle vv'\rangle$
that is a part of the boundary of both of them. A relation between the original
momentum variables $p_{\langle vv'\rangle}$ and the new variables $p_f$ is then given by:
\be\label{duality}
p_{\langle vv'\rangle}=p_f - p_{f'}.
\ee
Let us check that the number of new variables is the number of old variables minus the
number of constrains. The number of new variable is equal to $F-1$: the number of faces minus one.
We have to subtract one because everything depends on the differences only and does
not change if we shift all the variables by the same amount. We have $F-1=E-(V-1)-2G=E-(V-1)$,
where we have used our assumption that $G=0$. Thus, the number of new variables equals
to the number of the original variables, minus the number of $\delta$-functions,
equal to $V-1$ because one $\delta$-function is always redundant. In case $G\not=0$
one has to introduce an extra variable for each independent non-contractible loop
on the surface, whose number is $2G$. Thus, having expressed the momentum variables 
via $p_f$ using \eqref{duality}, we have automatically solved
all the conservation constraints. The Feynman amplitude becomes:
\be\label{dual-rep}
\int \prod_f dp_f \prod_{\langle ff'\rangle} G(p_f - p_{f'}).
\ee

We shall refer to \eqref{dual-rep} as the dual representation of a 2D Feynman
diagram. One can perform a similar duality transformation in higher dimensions.
The case relevant for us here is that of 3D, so let us briefly analyze what happens.
Let us consider a vacuum (closed) Feynman diagram and introduce an
additional structure that specifies faces, as well as an ordering of these faces around edges.
For example, let us consider Feynman graphs
whose edges and vertices are those of a {\it triangulation} of spacetime. As we
shall soon see, such Feynman diagrams are natural if one wants to interpret the
state sum models of 3D quantum gravity in particle terms. The faces of such 
``triangulation'' diagrams are then the usual triangles. As in 2D, let us assign
a variable $p_f$ to each face. The original edge momentum variable is then expressed
as an appropriate sum of the face variables, for all the faces that share the
given edge:
\be
p_e = \sum_{f: e\in f} p_f \epsilon_f.
\ee
One sums the face variables $p_f$ weighted with sign $\epsilon_f$; 
a precise convention is unimportant for us at the moment. The number of the
original variables is $E-(V-1)$, and due to the fact that the Euler characteristic
of any 3-manifold is equal to zero, we get: $E-(V-1)=F-(T-1)$, which means that
one in addition has to impose $T-1$ constraints for the new variables. 
Thus, in the original momentum representation we had $E$ edge variables
together with the momentum conservation constraints for each vertex.
In the dual formulation one has a momentum variable for each {\it dual edge},
and one conservation constraint for each dual vertex. We had to switch to the
dual lattice in order to solve the $\delta$-function constraints.

\subsection{Point particle Feynman diagrams}

Here we further develop our proposal of 
modifying Feynman diagrams by making the momentum group valued. Thus, let 
us consider some theory of point particles in 3D that generates Feynman
diagrams. A form of the action is unimportant for us. Our starting point
will be diagrams in the momentum representation. As we have already described, the most 
unusual feature of point particles in 3D is that their momentum is group-valued and thus non-commutative.
Thus, point particles in 3D behave unlike the standard relativistic fields in Minkowski spacetime.
In spite of this, one can still consider Feynman diagrams. Indeed, in the momentum
formulation \eqref{mom-rep} one only needs to specify what the propagator is, what the integration
measure $dp$ is, and what replaces the momentum conservation $\delta$-functions. All
this objects exist naturally for the 3D point particles. The propagator $G(p)$ is
some function on the group $\SU(2)$, which we shall leave unspecified for now. The
integration measure is the Haar measure on the group. The momentum conservation is 
more subtle. The conservation constraint becomes the 
condition that the product of the group elements $g_{\langle vv'\rangle}$ for all $v'$ 
is equal to the identity group element. However, it now does matter in which 
order the group elements are multiplied. Thus, some additional structure is
necessary to define the momentum conservation constraints and Feynman amplitude as
a whole. We shall specify this additional structure below. Ignoring this issue for the moment we get: 
\be\label{pp-ampl}
{\mathcal Z} = \int \prod_{\langle vv'\rangle} dg_{\langle vv'\rangle} G(g_{\langle vv'\rangle}) 
\prod_v \delta\left(\prod_{v'} g_{\langle vv'\rangle} \right).
\ee
Here $g_e=g_{\langle vv'\rangle}\in\SU(2)$ is the group element that describes particle's momentum on edge $e$,
$G(g)$ is a propagator which is a function on the group $\SU(2)$, and
$\delta(g)$ is the usual $\delta$-function on the group picked at the identity element.

Let us now deal with the issue of defining the momentum conservation constraints. It is
clear that some additional structure is necessary. We have already seen an extra structure
being added to a Feynman diagram when we considered the dual formulation. In that case
an extra structure was introduced to solve the momentum constraints. It is clear that
exactly the same structure can be used to define these constraints in the case of 
group-valued momenta. Thus, let us introduce an extra structure of faces, as well
as ordering of faces around each edge. This is equivalent to introducing the
structure of a dual complex. For simplicity we shall assume that the Feynman
diagram we started from is a triangulation of some manifold. The dual complex is then
the dual triangulation. Let us introduce new momentum variables: one for each face of the triangulation. The
original edge momentum is expressed as a product of the face variables:
\be\label{new-var}
g_e = \prod_{f: e\in f} g_f.
\ee
The order of the product here is important and is given by the ordering of the faces around the edge:
one multiples the holonomies across faces
to get the holonomy around the edge. The new face momentum variables $g_f$ solve
all the vertex momentum conservation constraints. 
The amplitude \eqref{pp-ampl} in the dual formulation becomes:
\be\label{pp-dual}
{\mathcal Z} = \int \prod_f dg_f \prod_e G\left(\prod_{f: e\in f} g_f\right) \prod_t TC,
\ee
where $TC$ are the tetrahedral constraints that were shown to be necessary by counting the
variables in the previous subsection. 

\subsection{Tetrahedron constraints}

It turns out that the tetrahedron constraints have a simple
meaning.\footnote{We thank Laurent Freidel for helping us with this
  interpretation.} The constraints imply that the face variables are
not all independent. There is a gauge symmetry that acts at each
tetrahedron. At this stage it is convenient to pass to the dual
triangulation. Recall that each face is dual to an edge $e$ of the dual
triangulation, and each tetrahedron is dual to a vertex $v$. To write
down the action of the gauge transformation acting at $v$, let us orient all the
edges $e$ incident at this vertex to point away from $v$. Then the
action on $g_e$ is as follows: $g_e\to g_v g_e$, where $g_v$ is the
same for all edges $e$ incident at $v$. Fixing this gauge symmetry is
the same as imposing the tetrahedron constraints in
\eqref{pp-dual}. Below we shall see how these constraints can be dealt with.

There is another possible interpretation that can be given to the constraints.
Let us consider a geometric tetrahedron in $\RR^3$.
The group valued face variables $g_f$ that we have introduced have the meaning of
an $\SU(2)$ transformation that has to be carried out when one goes from one tetrahedron
of the triangulation to the other. This group element can be represented as a product of
two group elements, one for each tet: $g_f = (g_f^t)^{-1} g_f^{t'}$, where $t,t'$ are the two
tetrahedra that share the face $f$. Consider now: $(g^t_f)^{-1} g^t_{f'}$. This group
element describes the rotation of one face $f$ into another $f'$, and thus 
carries information about the dihedral angle between the two faces. Thus, the group variables
$g_f$ introduced to solve the momentum constraints carry information about the dihedral
angles between the faces of the triangulation. In particular, the total angle $2\pi-\theta_e$ on an edge
$e$ is just the sum of all the dihedral angles around this edge:
\be
2\pi-\theta_e = \sum_{\langle ff'\rangle} \theta_{ff'},
\ee
where the sum is taken over the pairs $\langle ff'\rangle$ of faces that share the edge $e$.
This relation should be thought of as contained in the relation \eqref{new-var}.
Having understood the (partial) geometrical meaning of the face variables as encoding the dihedral
angles between the faces, we can state the (partial) meaning of the tetrahedron constraints. Recall
that 6 dihedral angles of a tetrahedron satisfy one relation. This relation is obtained as
follows. Consider the unit normals $\vec{n_i}, i=1,\ldots,4$ to all 4 faces of the tetrahedron.
Let us form a matrix $A_{ij}=\vec{n_i}\cdot\vec{n_j}$ of products of the normals. It is clear
that $\vec{n_i}\cdot\vec{n_j} = -\cos(\theta_{ij})$, and so the entries of $A_{ij}$ are
just the cosines of the dihedral angles. However, as the vectors $\vec{n_i}$ are 4 vectors in
$\RR^3$, they must be linearly dependent:
\be
\sum_i A_i \vec{n_i} = 0.
\ee
The coefficients $A_i$ here are just the areas of the corresponding faces. Because the normals
$\vec{n_i}$ are linearly dependent the determinant of the matrix $A_{ij}$ is equal to zero, 
which is the sought constraint among the 6 dihedral angles
$\theta_{ij}$. This relation between the dihedral angles should be
thought of as contained in the tetrahedron constraints in \eqref{pp-dual}.

Let us now consider some example theories to which the above strategy
can be applied.

\subsection{Pure gravity from point particle Feynman diagrams}

One special important case to be considered arises when particle's propagator is equal to the 
$\delta$-function of the momentum. Had we been dealing with a usual theory in $\RR^3$ in 
which the position and momentum representations are related by the Fourier transform
such a choice of the propagator would mean having $G(x-y)=1$ in the position space.
Thus, this theory is a trivial one as far as particles are concerned - there are no
particles. However, as we shall see in a moment, in the point particle context it
leads to a non-trivial and interesting amplitude. 

From the perspective of the Feynman amplitude \eqref{pp-ampl} it is clear that this case is a bit
singular. Indeed, the edge $\delta$-functions guarantee that all the holonomies
around edges are trivial. But then the vertex momentum conservation $\delta$-function
is redundant, and makes the amplitude divergent. To deal with this divergence, let
us consider a modified amplitude with no momentum conservation $\delta$-functions present. 
An equivalent way of dealing with this problem is to introduce the
(formal) quantity $\eta$ given by \eqref{eta}.
Now we have to divide our amplitude by $\eta^{-2V}$, where $V$
is the number of vertices in the diagram. We can now consider the dual formulation
in which all the constraints are solved. Let us keep the factors of $\eta$, for they
are going to be an important part of the answer that we are about to get. The amplitude
in the dual formulation is given by:
\be
{\mathcal Z} = \eta^{2V} \int \prod_f dg_f \prod_e \delta\left(\prod_{f: e\in f} g_f\right) \prod_t TC.
\ee
Now it is easy to verify that the expression under the integral is
invariant under the gauge transformations that act at the vertices of
the dual triangulation. Thus, the tetrahedron constraints are
redundant and can be dropped. This does not introduce any
divergences, as the extra integrations performed if one drops the
constraints gives the volume of the gauge group to the power of the
number of tetrahedra. We always assume that the Haar measure on the
group is normalized so that the volume of the group is one. Thus, when
a gauge invariant quantity is integrated, the tetrahedron constraints
can be dropped.

Thus, let us consider:
\be
{\mathcal Z} = \eta^{2V} \int \prod_f dg_f \prod_e \delta\left(\prod_{f: e\in f} g_f\right).
\ee
It is now possible to take all the group integrations explicitly. Indeed, 
decomposing all the $\delta$-function into characters, and performing the group integrations,
one gets exactly the Ponzano-Regge amplitude \eqref{PR}, with the important prefactor of
$\eta^{2V}$ that is missing in \eqref{PR-n}. It is only when this prefactor is present that 
the Ponzano-Regge amplitude is triangulation independent. 

Thus, we learn that the Ponzano-Regge amplitude \eqref{PR}, which, according to our previous 
discussion, is equal to the quantum gravity partition function, can also be interpreted
as an amplitude for a point particle Feynman diagram with the particle's propagator given by
the $\delta$-function of the momentum. A few comments are in order. First, the Ponzano-Regge
amplitude is independent of the triangulation chosen. Therefore, if we now take the particle interpretation, the 
Feynman amplitude is independent of a Feynman diagram. This could have been
expected, because particle's propagator given by the $\delta$-function essentially says that
no particle is present at all. This is why any Feynman diagram gives the same result.
The second comment is on why such two different
and seemingly contradictory interpretations are possible. Indeed, in one of the
pictures, we are dealing with the pure gravity partition function, and no particles
is ever mentioned. In the second picture, we consider point particles, whose momentum
became non-commutative group-valued. This happened due to the back-reaction of the
particle on the geometry. Point particle Feynman diagrams as integrals over
the group manifold take this back-reaction into account. Because pure gravity
does not have its own degrees of freedom, the only variables we have to integrate
over are those of the particles. Thus, particle's Feynman diagrams do give particle's
quantum amplitudes with quantum gravity effects taken into account. This is why no
separate path integration over gravity variables is necessary.

The picture emerging is very appealing. Indeed, one has two different interpretations
possible. In one of them one considers pure gravity with its ``topological'' degrees
of freedom. The other picture gives a ``materialistic'' interpretation of the pure
gravity result in terms of point particles. 

It is now clear that a generalization is
possible. Indeed, one does not have to take the propagator to be the $\delta$-function.
For example, one can consider the $\delta$-function $\delta_\theta$ picked on some non-trivial
conjugacy class $\theta$. The conjugacy classes can be chosen to be independent for
different edges. Then in the dual formulation, one arrives at the Ponzano-Regge
amplitude \eqref{PR-pp} modified by the presence of particles. 

The next level of generalization would be to allow all possible values of $\theta$ on 
every edge, and integrate over each $\theta_e$. This would mean integrating
over possible masses of point particles on every edge. Unlike the theory with a given
fixed set of angle deficits, which gives quantum gravity amplitudes modified by the presence
of a fixed configuration of point particles, theory that integrates over particle's
masses describes quantized particles together with quantized gravity. It is
our main goal in this paper to study what types of theories of this type are
possible. Thus, having understood that quantum gravity is described by 
taking into account a modification of particle's momentum that is due to the
back-reaction, the question to ask is what fixes the theory that describes
the particles. In other words, what is a principle that selects particle's
propagator? What fixes valency of the vertices, that is the type of
Feynman diagrams that are allowed? In the usual field theory we have the
action principle that provides us with an answer to all these questions. In our
case there is no such principle known. The models that we have considered so far
are only known in the momentum representation or the corresponding dual. 
In fact, because the momentum space is non-commutative
group manifold, one should expect the same to be true about the coordinate space. 
Thus, one can expect that the original coordinate representation of theories
of the type we consider is in some non-commutative space. It has been argued recently that 
even in the no-gravity limit $G\to 0$ this non-commutativity survives and leads to doubly special relativity, see
\cite{Freidel} as well as a more recent paper \cite{Majid-F} and references therein for more details.

Thus, one possible direction would be to understand what is the coordinate
representation for point particle theories, and then postulate some
action principle that would fix the propagator and interaction type. However, in this paper 
we shall instead consider the 
dual momentum formulation as fundamental. The question we address is which 
particle theories lead to ``natural'' dual representations. Indeed, to
arrive to the quantum gravity interpretation in terms of the Ponzano-Regge
amplitude we had to switch to the dual representation. It is the
dual representation that introduces a triangulation, and thus gives
some relation to geometry. Another argument in favor of the dual representation is 
that Feynman diagrams in the original momentum representation are insensitive to
the topology of the manifold they are on. Thus, they do not contain
all of the degrees of freedom of the gravitational field, only the
local point particle ones. To describe the global degrees of freedom
one needs to have the extra structure that is available in the dual 
formulation. Finally, even to define what the Feynman amplitude is in the case of a group-valued
momentum requires the dual representation. Indeed, to define the momentum conservation 
constraints one needs the structure of faces of the diagram, which is only
available with the dual graph. All this makes it rather clear that the dual
formulation is indispensable for the description of point particles in 3 dimensions. 
The logic of this paper will be to start from the formulation in terms of group field theory 
and the dual graph and then give the Feynman diagram interpretation.

With this idea in mind, we have to discuss how to obtain theories for
which a dual formulation is available. A natural way to do this is
using a generalization of the matrix model idea. Such a generalization
is available already for the Ponzano-Regge model, and is known
under the name of Boulatov theory \cite{Boulatov}. We shall describe
similar theories that incorporate point particles. Before we
do this, let us remind the reader some basic facts about the usual
matrix models.

\subsection{Matrix models}

Much of the activity in theoretical physics at the end of 1980's beginning of 1990's was 
concentrated in the area of matrix models. The idea is that the perturbative
expansion of a simple matrix model given by the action:
\be
S[M]= N {\rm Tr}\left( \frac{1}{2} M^2 + V(M) \right),
\ee
where $M$ is, say, a hermitian $N\times N$ matrix, and $V(M)$ is some (polynomial) potential,
can be interpreted as a sum over discretized random surfaces. Such a sum, in turn, gives
the path integral of Euclidean 2D gravity. The above matrix model can be solved
by a variety of techniques and the predicted critical exponents match those obtained
by the continuous methods. The simple one matrix model
given above corresponds to the theory of pure 2D gravity, or, if one chooses the potential
$V(M)$ appropriately, gives 2D gravity coupled to the so called $(2,2m-1)$ matter. Couplings to
other types of $c<1$ matter are possible by considering the multi-matrix models, see e.g. 
\cite{MM} and references therein for more details.

More generally, one can consider a matrix model for an infinite set of matrices parameterized
by a continuous parameter that is customarily referred to as time $t$. 
The corresponding matrix quantum mechanics 
gives 2D gravity coupled to a single scalar field, i.e. the so-called $c=1$ matter, 
and is given by the following action:
\be\label{mqm}
S[M(t)]=\int dt N {\rm Tr}\left( \frac{1}{2} \dot{M}^2+ \frac{m^2}{2} M^2 + V(M) \right).
\ee
Here the matrices $M(t)$ are functions of time, and $m$ is a ``mass'' parameter. 
The matrix quantum mechanics \eqref{mqm} can be solved exactly. The results can be compared with
those obtained in the continuous approaches, with full agreement, see \cite{MM} for more details.

\subsection{Field theory over a group manifold}

The idea of generating random discretized manifolds from a matrix model was generalized to 3 dimensions in
\cite{Boulatov}. Here one considers a field $\phi(g_1,g_2,g_3)$ 
on three copies of a group manifold $G$. The field $\phi(g_1,g_2,g_3)$ should in addition
satisfy an invariance property: $\phi(g_1g,g_2g,g_3g)=\phi(g_1,g_2,g_3)$.
One should think of each argument of the field as
a generalization of the matrix index in 2D. One considers the following simple action:
\be\label{GFT-1}
S[\phi] = \frac{1}{2} \int dg_1 dg_2 dg_3 \,\,
\phi^2(g_1,g_2,g_3) + \frac{g}{4!} \prod_{i<j} \int dg_{ij} \prod_i \phi(g_{ij},g_{ik},g_{il}).
\ee
Here the first product in the interaction term  is over pairs of indices $i,j=1,\ldots,4$, and 
indices $i,j,k,l$ in the arguments of $\phi$ are such that $i\not=j\not=k\not=l$.
Using the Fourier analysis on the group one can expand function $\phi$ into the group matrix elements.
One then obtains the Feynman rules from the action written in the ``momentum'' representation.
The resulting sum is over triangulated 3-manifolds, or, more
precisely, pseudo-manifolds. Edges of the Feynman graph correspond to faces
of the triangulation, and Feynman graph vertices 
corresponding to the tetrahedra. Edges of the complex get labeled by 
irreducible $G$-representations $l_e$ and one has to sum over these. The perturbative expansion
is given by:
\be
\sum_{complexes} g^T \sum_{\{l_e\}} \prod_e {\rm dim}(l_e) \prod_t (6j).
\ee
The first product here is over the edges of the triangulation, and the second is over its 3-cells 
(tetrahedra). The quantity $T$ is the total number of tetrahedra in a complex.
The quantity $(6j)$ is the Racah coefficient that depends on 6 representations.
The amplitude of each complex here is (almost, because of the missing
$\eta$ factors) the Ponzano-Regge amplitude \cite{PR}. 
We shall give details of the calculation that leads to this result in the next section.

Thus, the generalized matrix model of Boulatov \cite{Boulatov} gives a sum over random triangulated 3-manifolds
weighted by the exponent of the discrete version of the gravity action.  Unlike the case of 2D matrix models
it was not possible to solve the Boulatov theory, and not so much has been learned from it about
3D quantum gravity, see, however, \cite{Sum} for an interesting proposal for summing over the
3-topologies. Similar theories have been considered in higher number of
dimensions as well. This more general class of models defined on the group manifold
and leading to sums over triangulated manifolds have received the name of group field
theories (GFT). 
  
\subsection{Outline and organization of the paper}

We have seen that the
Ponzano-Regge amplitude has a point particle interpretation: it is the amplitude 
for a theory of point particles with a propagator being the
$\delta$-function in the momentum space, and the amplitude is computed
in the dual momentum representation. In this paper we shall address the following question: 
what kind of group field theory can give dual momentum representation for point particle theory 
with a non-trivial propagator? As we shall see, the requirement that Feynman diagrams 
generated by group field theory also admit an interpretation as dual of some other
Feynman diagrams severely restricts the possible choices of theories. There is essentially
a few possible choices, which we shall discuss in due course. 

The main idea of this paper is to consider a group field theory on the {\it Drinfeld double}
$\D$ of the group $\SU(2)$. To the best of our knowledge, the fact that point particles in 3D are 
naturally described by $\D$ was discovered and explored in \cite{BM}. More recently, it
was used in \cite{Freidel-CS} to give the CS formulation of the Ponzano-Regge
model. As we shall see, a group field theory construction on $\D$ similar
to that of Boulatov \cite{Boulatov} can be performed. In some sense, the
idea of going from $\SU(2)$ group field theory to $\D$ one is analogous to
the way one introduces the matter degrees of freedom in 2D, where multi-matrix instead
of one-matrix models are considered. However, the analogy here is not
direct, for, as we shall find, the most obvious generalization does
not lead to a theory with point particles in it. Nevertheless, group
field theory considerations will allow us to develop certain methods
that will later be used to describe a general point particle theory.

The organization of this paper is as follows. In the next section we
describe how Ponzano-Regge model is obtained from Boulatov
theory. Here we also give an abstract algebraic formulation of
Boulatov model that will be later used to define GFT for the Drinfeld
double. We describe the Drinfeld double of a classical group in
section \ref{sec:DD}. Here we discuss irreducible representations of
$\D$, characters, as well as an important projector property satisfied
by them. Group field theory for the Drinfeld double is defined in
section \ref{sec:gft}, and some of its properties are proved. Section
\ref{pp-models} defines more general particle models that are not
related to GFT. Interpretation of all the models is developed
in section \ref{sec:pp}. We conclude
with a discussion of the results obtained.

Let us note that, apart from the works already mentioned, the question of coupling of quantum 
gravity to matter was considered in paper \cite{Mikovic}. However, the approach taken in the 
present paper is rather different.

\section{Boulatov field theory over the group}
\label{sec:Boul}

Boulatov theory \cite{Boulatov} can be formulated in several equivalent ways. One of
this formulation will be used for generalization to the Drinfeld double. Let us
first give the original formulation used in \cite{Boulatov}

\subsection{Original Boulatov formulation}

Let us consider a scalar field on the group manifold $\phi(g)$. For definiteness we take the group to
be $G=\SU(2)$ that corresponds to 3D Euclidean gravity. 
However, one can take any (compact) Lie group. The resulting theory would still be topological, and be
related to the so-called BF theory for the group $G$. The
Fourier decomposition of the function $\phi$ is given by:
\be\label{b1}
\phi(g) = \sum_{l} \sum_{mn} t^l_{mn}(g) \phi^l_{mn}.
\ee
Here $t^l_{mn}(g)$ are the matrix elements in $l$'th representation, and $\phi^l_{mn}$ are the
Fourier coefficients. It is extremely helpful to introduce a graphic notation for this formula. 
Denoting the Fourier coefficients by a box, and the matrix elements by a circle, both with two lines 
for indices $m,n$ sticking out, we get:
\be
\phi(g) = \sum_l \,\,\, \lower0.23in\hbox{\epsfig{figure=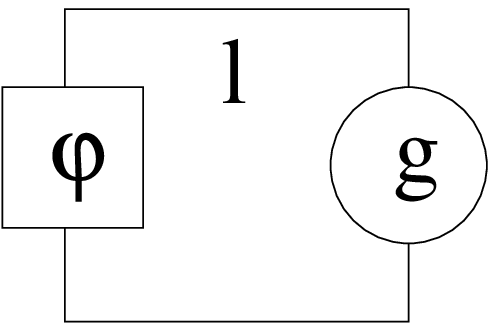, height=0.45in}}
\ee
The sum over $m,n$ is implied here. The graphical notation is much
easier to read than \eqref{b1}, and we shall write many formulas using
it in what follows.

The field of Boulatov theory is on 3 copies of the group manifold. Analogous Fourier expansion is
given by:
\be
\phi(g_1,g_2,g_3) = \sum_{l_1,l_2,l_3} \,\,\, \lower0.25in\hbox{\epsfig{figure=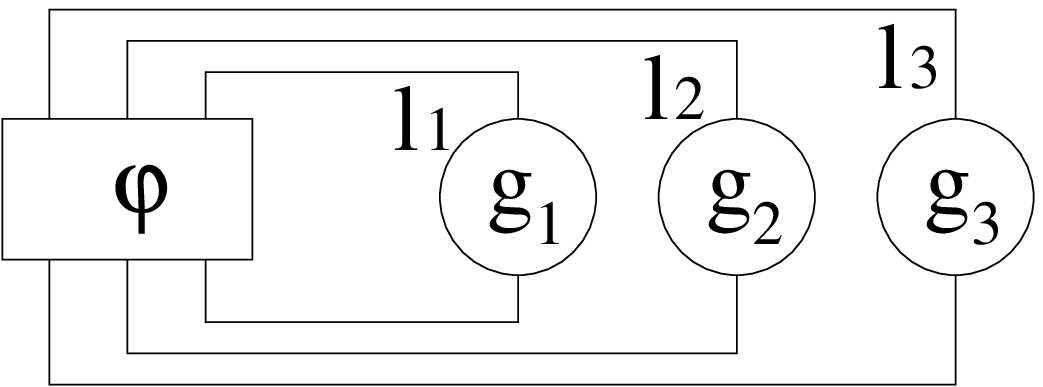, height=0.5in}}
\ee
The field is required to be symmetric:
\be
\phi(g_{\sigma(1)},g_{\sigma(2)},g_{\sigma(3)})=(-1)^{|\sigma|} \phi(g_1,g_2,g_3).
\ee
Here $\sigma$ is a permutation, and $|\sigma|$ its signature.
In addition, it is required to satisfy the following invariance property:
\be\label{sym}
\phi(g_1 g,g_2 g,g_3 g)=\phi(g_1,g_2,g_3).
\ee
Let us find consequences of this for the Fourier decomposition. Let us integrate the right hand side of
\eqref{sym} over the group. We are using the normalized Haar measure, so we will get $\phi(g_1,g_2,g_3)$
on the right hand side. On the left hand side we can use the formula \eqref{A-3} for the integral of three
matrix elements. Thus, we get:
\be\label{phi-1}
\phi(g_1,g_2,g_3) = \sum_{l_1,l_2,l_3} \,\,\,\lower0.5in\hbox{\epsfig{figure=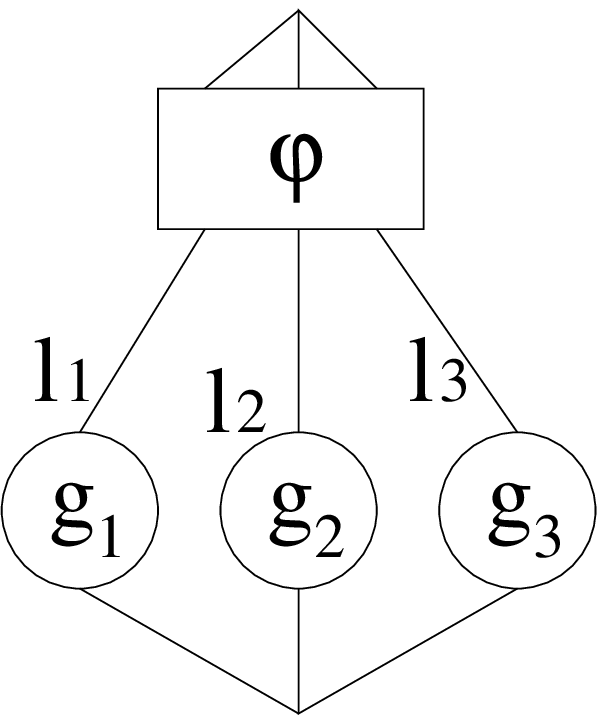, height=1in}}
\ee
Let us now introduce a new set of Fourier coefficients $\tilde{\phi}^{l_1 l_2 l_3}_{m_1 m_2 m_3}$.
Graphically, they are defined by:
\be\label{tilde-phi}
\lower0.25in\hbox{\epsfig{figure=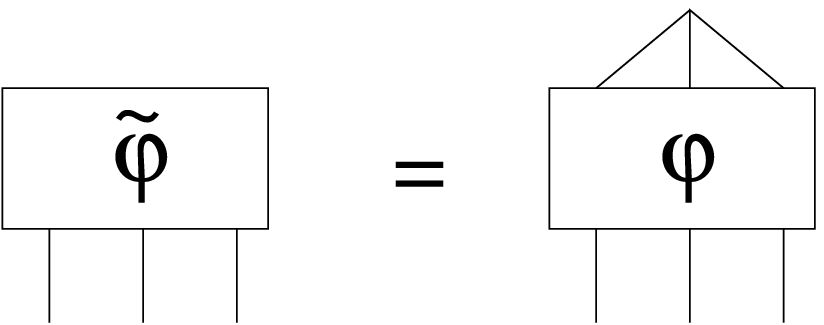, height=0.5in}}
\ee
We shall only use the new, modified set of coefficients
\eqref{tilde-phi} till the end of this section. Thus, we
shall omit the tilde. Our final expression for the Fourier decomposition is given by:
\be\label{phi-2}
\phi(g_1,g_2,g_3) = \sum_{l_1,l_2,l_3} \,\,\,\lower0.45in\hbox{\epsfig{figure=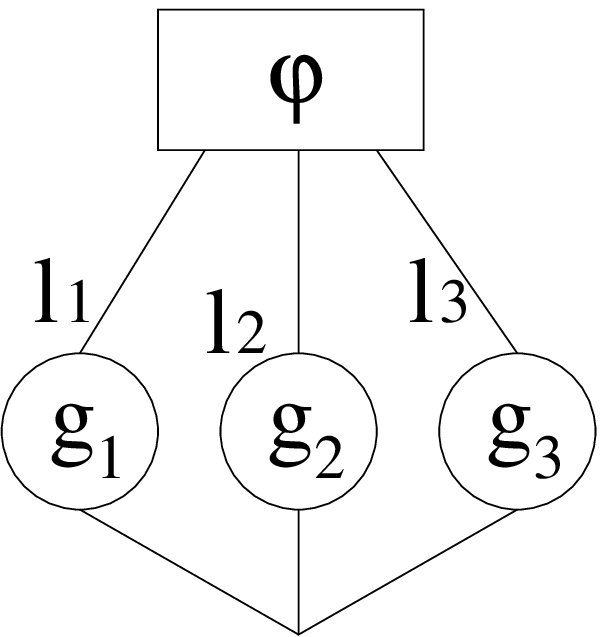, height=0.9in}}
\ee

It is now straightforward to write an expression for the action in terms of the Fourier coefficients. 
The action \eqref{GFT-1} is designed in such a way that there is always two matrix elements containing
the same argument. Thus, a repeated usage of the formula \eqref{A-2} gives:
\be\label{action-f}
S[\phi] = \frac{1}{2} \sum_{l_1 l_2 l_3}\!{}' \,\,\,\lower0.25in\hbox{\epsfig{figure=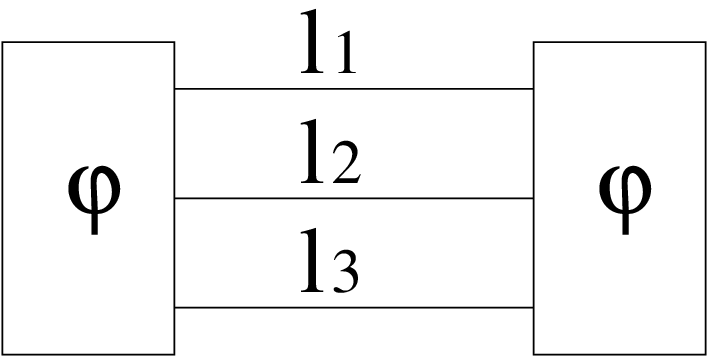, height=0.5in}}
+ \\ \nonumber
\frac{g}{4!} \sum_{l_{ij}}\,\,\,\left( \begin{array}{ccc} l_{12} & l_{13} & l_{14} \\
l_{34} & l_{24} & l_{23} \end{array}\right) \,\,\, \lower0.5in\hbox{\epsfig{figure=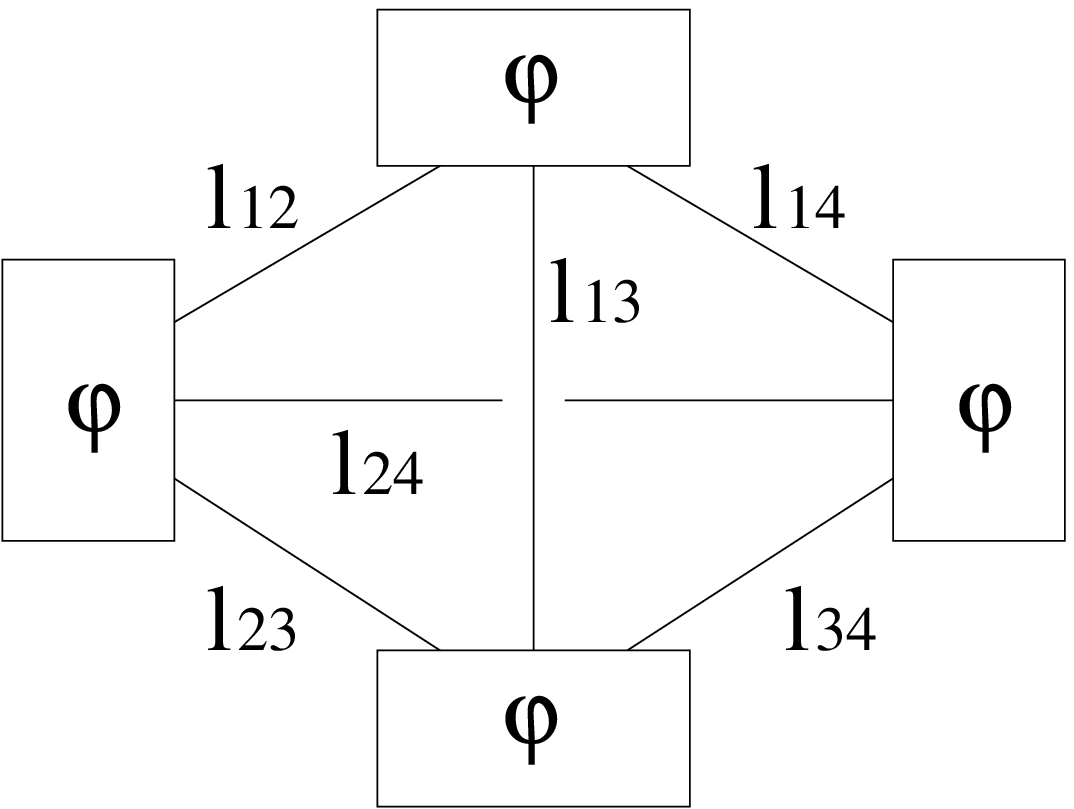, height=1in}}
\ee
Here the first sum is taken only over sets $l_1,l_2,l_3$ that satisfy
the triangular inequalities, and this is reflected in a prime next to the sum symbol.
To arrive to this expression we have used the normalization of the intertwiner given by \eqref{A-theta}.
We have also introduced the so-called Racah coefficient, or the $6j$-symbol that is denoted by
two rows of spins in brackets. 

It is now easy to derive the Feynman rules. The propagator and the vertex are given by, correspondingly:
\be
\lower0.23in\hbox{\epsfig{figure=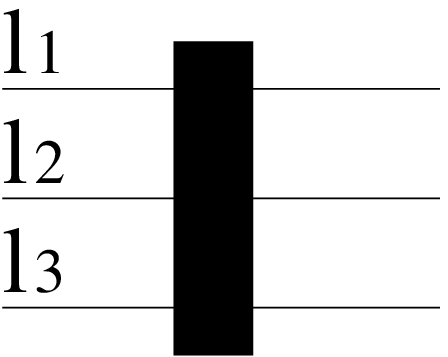, height=0.45in}} 
\ee
where the black box denotes the sum over permutations, and 
\be
g \left( \begin{array}{ccc} l_{12} & l_{13} & l_{14} \\
l_{34} & l_{24} & l_{23} \end{array}\right)
\,\,\, \lower0.45in\hbox{\epsfig{figure=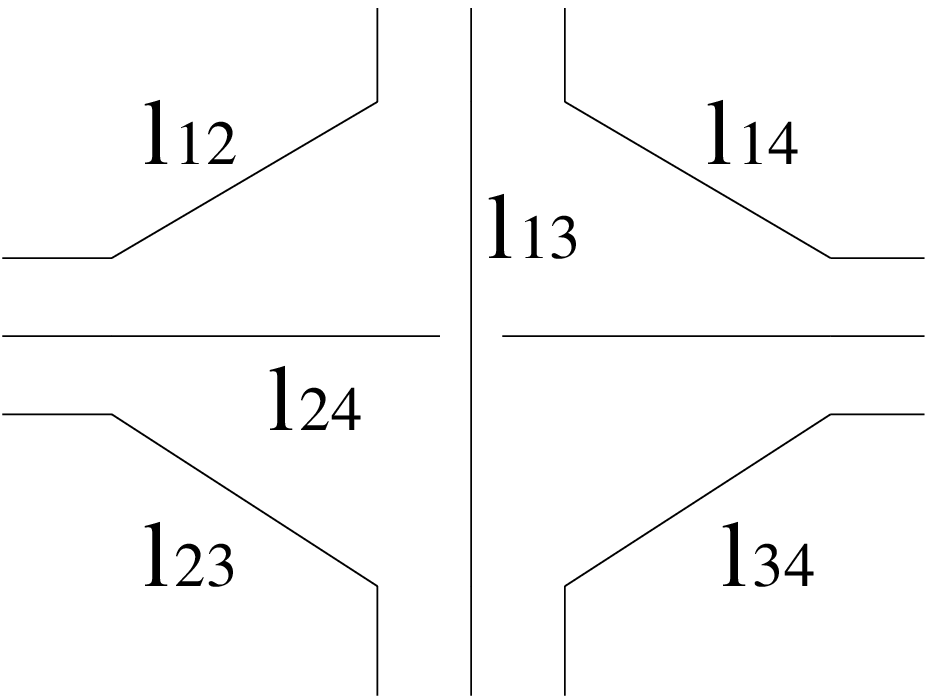, height=0.9in}}
\ee
Note now that each Feynman graph of this theory is a dual skeleton of some simplicial 3D manifold. Indeed,
vertices of the graph correspond to simplices, edges corresponds to faces, and faces 
(closed loops) corresponds to edges. A proper way to 
describe the 3-manifolds arising is in terms of pseudo-manifolds. It will be presented below.
Each Feynman amplitude is weighted by {\it almost} the Ponzano-Regge amplitude.
What is missing is a (vanishing for the classical group) factor of $\eta^2$ for each vertex of the triangulation.

Note that, unlike the
case of 2D gravity where we have a clear interpretation of both the rank $N$ of the matrices
and the coupling constant $g$ (they are related to Newton and cosmological constant correspondingly), 
in the case of Boulatov theory the interpretation of $g$ is obscure.
It is not anymore related to the volume of a tetrahedron, for the latter is now a function of the
spins. The appearance of Newton's constant is also not that direct. It serves to relate the
spin $l_e$ labeling edges to their physical (dimensionful) length. It has been argued recently \cite{Laurent-recent}
that the coupling constant $g$ should be thought of as the loop counting parameter that weights the
topology changing processes. However, this issue is not settled and we shall not comment on it any further.

\subsection{Algebra structure}
\label{ss:algebra}

Let us remind the reader that the algebra ${\mathcal A}^*$ of functions on
the group can be given a structure of the Hopf algebra. The reason why the algebra is referred to as
${\mathcal A}^*$ and not $\mathcal A$ will become clear
below.\footnote{It is customary to denote the commutative point-wise
  product of functions by $\bullet$ and the non-commutative
  convolution product by $\star$, so we go against the convention
  here. The reason for this unusual notation is that we decided to
  embed $\mathcal A$ into the Drinfeld double $\D$ in a particular
  way, see below, and this induces the $\bullet$ and $\star$ products
  in the way chosen.} The Hopf algebra structure is as follows: 
\be\nonumber
{\rm multiplication} &{}& (f_1\star f_2)(x)= f_1(x) f_2(x), \\ \nonumber
{\rm identity} &{}& i(x)= 1, \\ \nonumber
{\rm co-multiplication} &{}& (\Delta^* f)(x,y) = f(xy), \\
{\rm co-unit} &{}& \varepsilon(f) = f(e),  \\ \nonumber
{\rm antipode} &{}& (\kappa f)(x)=f(x^{-1}), \\ \nonumber
{\rm involution} &{}& f^\circ(x) = \overline{f(x)}.
\ee
Of importance is also the Haar functional $h:{\mathcal A}\to \CC$:
\be
h^*(f) = \int dx\, f(x).
\ee
A non-degenerate pairing
\be
\langle f_1,f_2 \rangle = \int dx f_1(x) f_2(x)
\ee
identifies ${\mathcal A}^*$ with its dual $({\mathcal A}^*)^*={\mathcal A}$. 

The dual algebra $\mathcal A$ is also a Hopf algebra. A multiplication on $\mathcal A$ is denoted by $\bullet$ and is
introduced via:
\be
\langle f_1 \bullet f_2, g\rangle := \langle f_1 \otimes f_2, \Delta^* g\rangle.
\ee
One obtains:
\be\label{bullet}
(f_1\bullet f_2)(x) = \int dz\, f_1(z) f_2(z^{-1} x).
\ee
All other operations on $\mathcal A$ read:
\be
{\rm identity} &{}& i(x)= \delta_e(x), \\ \nonumber
{\rm co-multiplication} &{}& (\Delta f)(x,y) = f(x) \delta_x(y), \\
{\rm co-unit} &{}& \varepsilon(f) = \int dx f(x),  \\ \nonumber
{\rm antipode} &{}& (\kappa f)(x)=f(x^{-1}), \\ \nonumber
{\rm involution} &{}& f^*(x) = \overline{f(x^{-1})}.
\ee
It is easy to check that $*$ is indeed an involution:
\be\label{inv}
(f_1\bullet f_2)^* = f_2^* \bullet f_1^*.
\ee
One also needs a Haar functional $h:{\mathcal A}\to \CC$:
\be
h(f)=f(e).
\ee

Using this structure, a positive definite inner product can be
defined:
\be\label{ip}
\langle f_1, \overline{f_2} \rangle = \int dx f_1(x) \overline{f_2(x)}
= h(f_1 \bullet f_2^*)=h(f_2^* \bullet f_1).
\ee
The algebraic structure on $\mathcal A$ will play an important role when we
give an algebraic formulation of the Boulatov theory.

\subsection{Projectors}

Of special importance are functions on the group that satisfy the
projector property:
\be\label{proj-prop}
f\bullet f = f.
\ee
Examples of such projectors are given by:
(i) characters 
\be\label{P-j}
P_j:={\rm dim}_j \chi_j(x),
\ee
and (ii) spherical functions ${\rm
  dim}_j T^j_{00}(x)$. In both cases the projector property is readily
verified using \eqref{A-2-1}. We have:
\be
P_j \bullet P_k = \delta_{jk} P_j.
\ee
We will mostly be interested in these
character projectors. We note that the unit element, that is, the 
$\delta$-function on the group
can be decomposed into the character projectors:
\be
i(x)=\delta_e(x)=\sum_j P_j.
\ee
Of importance for what follows is the following element of ${\mathcal A}^*\otimes{\mathcal A}^*$:
\be\label{G-prop}
G=(\kappa\otimes {\rm id})\Delta^* i, \\ \nonumber
G(x,y) = \delta_e(x^{-1} y).
\ee
In view of a property: 
\be
\int dy\, G(x,y) G(y,z) = G(x,z)
\ee
this function on $G\times G$ can be referred to as a {\it propagator}. The propagator is just a kernel 
of the operator $G$ acting on $\mathcal A$ via the $\bullet$-product.

The unit element and the characters are the basic projectors on a
single copy of $\mathcal A$. More interesting projectors
can be constructed when working with several copies of the algebra. 
Consider an element $\theta\in {\mathcal A}^{\otimes 3}$ obtained from 
three propagators $G$:
\be
\theta:= (1^{\otimes 3} \otimes h_3) G\otimes G \otimes G.
\ee
Here we have introduced a new operation $h_3:{\mathcal A}^* \otimes {\mathcal A}^* \otimes {\mathcal A}^* \to\CC$:
\be
h_3(f_1,f_2,f_3) = h^*(f_1\star f_2\star f_3) = \int dx f_1(x) f_2(x) f_3(x).
\ee
As a function on 3 copies of the group the element $\theta$ is given by:
\be\label{theta}
\theta(x_1,x_2,x_3) := \int dg\, \delta_e(x_1^{-1} g) \delta_e(x_2^{-1} g)
\delta_e(x_3^{-1} g) = 
\sum_{j_i} \prod_i{\rm dim}_{j_i} \theta_{j_i}(x_i),
\ee
where
\be\label{theta-j}
\theta_{j_i}(x_i):= \theta_{j_1,j_2,j_3}(x_1,x_2,x_3):=\int dg\,\, \chi_{j_1}(x_1^{-1} g) \chi_{j_2}(x_2^{-1} g) \chi_{j_2}(x_2^{-1} g).
\ee 
Let us note that $\theta$ is $*$-invariant: $\theta^*=\theta$.
It is easy to verify that $\theta$ is a projector:
\be
\theta\bullet \theta = \theta.
\ee
To show this one uses \eqref{A-3} and \eqref{A-theta}. The function
$\theta(x_1,x_2,x_3)$ projects onto functions invariant under the left
diagonal action. Indeed, consider an arbitrary function
$\phi(x_1,x_2,x_3)\in {\mathcal A}^{\otimes 3}$. Define
the $\Phi$ to be $\theta$ with the projector $\theta$ applied on the
left. The function $\Phi$ is invariant under the left diagonal shifts:
\be
\tilde{\phi}:=\theta\bullet \phi, \qquad \tilde{\phi}(g x_1,g x_2, g
x_3)=\tilde{\phi}(x_1,x_2,x_3).
\ee
The field $\tilde{\phi}$ obtained this way is the basic field of Boulatov
theory, see \eqref{sym}. The above more abstract formulation in
algebra terms in necessary for a generalization to field theory
on the Drinfeld double.

\subsection{Algebraic formulation of Boulatov's theory}

An equivalent formulation of Boulatov theory can be 
achieved using the Hopf algebra operations introduced 
earlier in this section. More specifically, we will need
the second algebra structure in terms of the non-commutative
$\bullet$-product and $*$-involution. Let us first write the
kinetic term of the field theory action \eqref{GFT-1}. We have:
\be\label{alg-kin}
S_{kin}=\frac{1}{2}h((\theta\bullet\phi)^* \bullet (\theta\bullet\phi)) =
\frac{1}{2}h(\phi^* \bullet\theta\bullet\phi).
\ee
Here we have used the fact \eqref{inv} that $*$ is an involution,
and the fact that $\theta$ is a projector. All operations here, namely 
the $\bullet$-product, the $*$-involution, and the co-unit are
understood in this formula as acting on ${\mathcal A}^{\otimes 3}$, 
separately on each of the copies of the algebra.

There are two possible points of view on the kinetic term
\eqref{alg-kin}. One is that the action is a functional of a field
$\tilde{\phi}$ that satisfies the invariance property
$\theta\bullet\tilde{\phi}=\tilde{\phi}$. This is the point of view
taken in the original formulation of Boulatov theory that has been
described above. The other interpretation is suggested by
\eqref{alg-kin}, and is to view $\theta$ in this expression as the
kinetic term ``differential operator'', or the inverse of the
propagator of the theory. The action is then invariant under the
following symmetry: $\delta\phi=\psi$, where $\psi$ is a field
satisfying $\theta\bullet\psi=0$. This symmetry should be viewed as
gauge symmetry of the theory; the ``physical'' degrees of freedom are
not those of $\phi$ but those of the $\theta$-cohomology classes.  
Boulatov theory in its original formulation \eqref{GFT-1}
can be viewed as the gauge-fixed version of this theory. The gauge symmetry
described is of great importance. For example, one can consider a theory
with no $\theta$-projector inserted in the 
action. This theory does not have any gravitational interpretation, as it is
easy to check. Thus, it is the presence of $\theta$ in the action, and thus
the extra gauge symmetry that ensures a relation to gravity. It can be 
argued that this symmetry is the usual diffeomorphism-invariance
of a gravitational theory in disguise. 

Once the kinetic term is understood, it is easy to write down the
interaction term as well. One has to take 4 fields $\theta\bullet\phi$
and $\bullet$-multiply them all as is suggested by the structure of the
interaction term in \eqref{GFT-1} to obtain an element of 
${\mathcal A}^{\otimes 6}$. Each copy of the algebra is a product of
two fields; the $*$-involution has to taken on one of them. After
all fields are multiplied, the Haar functional $h$ has to be applied to get a
number. We will not write the corresponding expression as it is rather
cumbersome. The structure arising is best understood using the language of operator kernels
to which we now turn.

\subsection{Formulation in terms of kernels}

We have given an algebraic formulation of Boulatov theory, in which fields are viewed 
as operators in ${\mathcal A}^{\otimes 3}$, and one uses the
$\bullet$-product on $\mathcal A$ to write the action. An equivalent formulation can be given by introducing
kernels of all the operators. Such a formulation is more familiar in
a field theory context, and
will be quite instrumental in dealing with the theory. 

Let us interpret the formula \eqref{bullet} as follows. The quantity: 
\be
f_2(x,y):=f_2(x^{-1} y) = (\kappa\otimes {\rm id})\Delta^* f_2, 
\ee
which is an element of ${\mathcal A}^*\otimes {\mathcal A}^*$,
is interpreted as the kernel of an operator ${\mathcal O}_{f_2}$ corresponding to $f_2$ that acts on $f_1$ from 
the right:
\be
f_1 \circ {\mathcal O}_{f_2} = \int dz f_1(z) f_2(z,x).
\ee
Thus, instead of dealing with operators from $\mathcal A$ one can work with kernels from
${\mathcal A}^*\otimes {\mathcal A}^*$. 

The inner product \eqref{ip} can also be expressed in terms of the kernels. Since
$f(y,z):=f(y^{-1} z)$, the kernel for $f^*$ is given by:
\be
f^*(y,z)=\overline{f(y z^{-1})}.
\ee
Therefore:
\be
\langle f_1, \overline{f_2}\rangle = \int dx dy \,\, f_1(x,y) f_2^*(y,x).
\ee
Here we have taken a convolution of the kernels of two operators, and then took the
Haar functional given by the trace of the corresponding kernel:
\be
h(f) \to \int dx f(x,x).
\ee

Let us now introduce kernels for all the objects that are necessary to define Boulatov theory. The 
kernel for the field $\phi\in {\mathcal A}^*\otimes {\mathcal A}^*\otimes {\mathcal A}^*$ is:
\be\label{ker-phi}
\phi := ((\kappa\otimes {\rm id})\Delta^*)^{\otimes 3} \phi, \\ \nonumber
\phi(x_1,y_1;x_2,y_2;x_3,y_3):= \phi(x_1^{-1} y_1,x_2^{-1} y_2, x_3^{-1} y_3).
\ee
The kernel for the projector $\theta$ is similarly given by:
\be\label{ker-theta}
\theta(x_1,y_1;x_2,y_2;x_3,y_3):= \theta(x_1^{-1} y_1,x_2^{-1} y_2, x_3^{-1} y_3).
\ee
The action of $\theta$ on the field now reads:
\be
&{}&\Phi(x_1,y_1;x_2,y_2;x_3,y_3):= \\ \nonumber
&{}& \int dz_1 dz_2 dz_3 \, \phi(x_1,z_1;x_2,z_2;x_3,z_3) 
\theta(z_1,y_1;z_2,y_2;z_3,y_3).
\ee

It is now easy to write the action for the theory. We shall assume that the
field is real. The action reads:
\be\label{boul-kern}
S= \frac{1}{2} \int \prod_{i=1}^3 dx_i dy_i dz_i\,\, \phi(x_i,y_i) \theta(y_i,z_i) \phi(z_i,x_i) +
\\ \nonumber
\frac{g}{4!} \int \prod_{i\not=j=1}^{4} dx_{ij} dy_{ij} dz_{ij} \,\, 
\phi(x_{ij},y_{ij}) \theta(y_{ij},z_{ij}) \theta(z_{ji},y_{ji}) \phi(y_{ji},x_{ji}).
\ee
Here $x_{ij}=x_{ji}, z_{ij}=z_{ji}$, but $y_{ij}\not=y_{ji}$ and are independent
integration variables. We have also introduced a notation: $\phi(x_i,y_i):=\phi(x_1,y_1;x_2,y_2;x_3,y_3)$ 
and $\phi(x_{ij},y_{ij}):=\phi(x_{ij},y_{ij};x_{ik},y_{ik};x_{il},y_{il})$ with
$i\not=j\not=k\not=l$.

\subsection{Remarks on Boulatov theory}

Let us remark on possible choices of propagators in
\eqref{boul-kern}. One can try to use a different operator in place of $\theta$ in both 
the kinetic and the potential terms. Or, more generally, one can have an 
interaction term that is built of quantities different than the one used
in the kinetic term, see \cite{Rovelli} where such more general theories
are considered. However, this more general class of theories fails to
lead to 3-manifold invariants. Related to this is the fact that these
more general theories do not have a dual Feynman diagram
interpretation. As we have discussed in the introduction, we would
like to restrict our attention to a special class of theories, namely
those in which Feynman amplitudes that follow from the group field
theory expansion also have an interpretation in terms of Feynman
diagrams for the dual complex. In other words, as we have discussed,
group field theory Feynman diagrams have the interpretation of
amplitudes for a complex dual to some triangulated 3-manifold. The
triangulation itself can be considered as a Feynman diagram. When can
the group field theory amplitude, which is the one for the dual triangulation, 
be interpreted as a Feynman amplitude for the original triangulation
as well? Thus, the question we are posing is which group field theories
admit a dual formulation. As is clear from our discussion of Boulatov and
Ponzano-Regge models in the introduction, this particular theory does admit
both interpretations. Are there any other theories with a similar property?

We shall not attempt to answer this question in its full generality. We shall only
make a remark concerning the choice of the propagator of the model. As we have seen,
the propagator of any 3d group field theory model should consist of 3 strands. Thus,
it can be described as made out of 3 propagators, one for each strand. This is
what happens in the case of Boulatov model, where the $\theta$-projector is made
out of 3 $G$-propagators \eqref{G-prop}. Now each strand of the group field
theory diagram that forms a closed loop is interpreted as dual to an
edge of a triangulated 3-manifold.
When this triangulation is itself interpreted as a Feynman diagram, there
must be a propagator for every edge. It is clear that this propagator will be
built from the group field theory strand propagator $G$, and will just be
a certain power of it, with the power given by a number of dual edges
forming the boundary of the dual face. Because different triangulations
can have this number different, to have the interpretation we are after
the propagator $G$ must be a projector $G\bullet G=G$. Only such 
propagators admit the dual Feynman diagram interpretation. This requirement
is clearly very restrictive, and limits the choice of possible propagators
dramatically. In this sense the models of the type we are considering in this paper
are very scarce.

Now that we have understood how to formulate Boulatov's theory in abstract terms, let us
construct an analogous theory with the group manifold replaced by the Drinfeld double of
$\SU(2)$.

\section{Drinfeld double of $\SU(2)$}
\label{sec:DD}

The Drinfeld double of a classical group was first studied in \cite{Philippe}. 
Our description of the Drinfeld double $D(G)$ of $G=\SU(2)$ closely
follows that in \cite{Koorn}. Following this reference, we describe 
$D(G)$ as the space of functions on it. As a linear space $D(G)$ is identified with the space
$C(G\times G)$ of functions on two copies of the group. On $D(G)$ we have a non-degenerate pairing:
\be
\langle f_1,f_2 \rangle := \int dx dy\,\, f_1(x,y) f_2(x,y).
\ee
This pairing identifies the dual $D(G)^\star$ of the Drinfeld double with $C(G\times G)$.
The following operations are defined on $D(G)$:
\be
\nonumber
{\rm multiplication} &{}& (f_1\bullet f_2)(x,y)= \int dz\, f_1(x,z) f_2(z^{-1}xz,z^{-1}y), \\ \nonumber
{\rm identity} &{}& 1(x,y)= \delta_e(y), \\ \nonumber
{\rm co-multiplication} &{}& (\Delta f)(x_1,y_1;x_2,y_2) =
f(x_1x_2,y_1)\delta_{y_1}(y_2), \\ \label{dd}
{\rm co-unit} &{}& \varepsilon(f) = \int dy\, f(e,y),  \\ \nonumber
{\rm antipode} &{}& (\kappa f)(x,y)=f(y^{-1} x^{-1} y,y^{-1}), \\ \nonumber
{\rm involution} &{}& f^\ast(x,y) = \overline{f(y^{-1} xy,y^{-1})}.
\ee

By duality we have the following operations on the dual $D(G)^*$:
\be
\nonumber
{\rm multiplication} &{}& (f_1\star f_2)(x,y)= \int dz\, f_1(z,y) f_2(z^{-1}x,y), \\ \nonumber
{\rm identity} &{}& i(x,y)= \delta_e(x), \\ \nonumber
{\rm co-multiplication} &{}& (\Delta^\star f)(x_1,y_1;x_2,y_2) =
f(x_1,y_1y_2)\delta_{x_2}(y_1^{-1}x_1y_1), \\ \label{dd-dual}
{\rm co-unit} &{}& \varepsilon^\star(f) = \int dx\, f(x,e),  \\ \nonumber
{\rm antipode} &{}& (\kappa^\star f)(x,y)=f(y^{-1} x^{-1} y,y^{-1}), \\ \nonumber
{\rm involution} &{}& f^\circ(x,y) = \overline{f(x^{-1},y)}.
\ee

The universal R-matrix $R\in D(G)\otimes D(G)$ is given by:
\be\label{R-matr}
R(x_1,y_1;x_2,y_2)=\delta_e(y_1) \delta_e(x_1 y_2^{-1}).
\ee
We will also need the central ribbon element:
\be
c(x,y)=\delta_e(xy)
\ee
and the monodromy element:
\be
Q(x_1,y_1;x_2,y_2):=(R\bullet \sigma R)(x_1,y_1;x_2,y_2)=\delta_{y_1}(x_2) \delta_{y_2}(x_2^{-1}x_1 x_2).
\ee
Here $\sigma(x_1,y_1;x_2,y_2)=(x_2,y_2;x_1,y_1)$.

The Haar functionals $h:D(G)\to \CC$ and $h^\star:D(G)^*\to\CC$ are given by:
\be
h(f)=\int dx\, f(x,e), \qquad h^\star(f)=\int dy\, f(e,y).
\ee
Note that the Haar functional $h^*$ on $D(G)^*$ coincides with the
co-unit $\epsilon$ on $D(G)$, and similarly $h$ coincides with $\epsilon^*$.
Using these functionals, a positive-definite inner product on $D(G)$ can be
defined as:
\be
\langle f_1, \overline{f_2} \rangle = \int dx dy\, f_1(x,y)
\overline{f_2(x,y)} = h(f_1\bullet f_2^*) = h^*(f_1 \star f_2^\circ).
\ee

To acquire a better understanding of the Drinfeld double, let us
consider some of its sub-algebras.

\subsection{Sub-algebras of the Drinfeld double}

The Drinfeld double is a twisted product of the algebra of functions on the
group and the group itself.
The subalgebra ${\mathcal A}^*$ of functions on the group described in the previous 
section is represented by the elements:
\be
f(g) \to \delta_e(x) f(y) \in D(G).
\ee
Then it is easy to check that the $\star$-product coincides with the usual point-wise
multiplication of functions: 
\be
(\delta_e f_1 \star \delta_e f_2)(x,y)=\delta_e(x) f_1(y) f_2(y).
\ee
It is therefore clear that the $\star$-algebraic structure on the algebra ${\mathcal A}^*$ of functions
on the group that we have described in section \ref{sec:Boul} is
exactly the one obtained from the Drinfeld double dual algebra structure
\eqref{dd-dual} when specialized to functions of the form $\delta_e(x) f(y)$. 
Note that the algebra ${\mathcal A}^*$
with its pointwise multiplication can also be embedded into the Drinfeld algebra itself
with its $\bullet$-product. Indeed, as is easy to check, the $\bullet$-product on
functions of the form $f(x) \delta_e(y)$ is just the usual pointwise multiplication.

The group is represented by elements of the form:
\be
g\to \delta_g(y) \in D(G).
\ee
The $\bullet$-multiplication reduces to the usual group multiplication law:
\be
(\delta_{g_1} \bullet \delta_{g_2})(x,y) = \delta_{g_1g_2}(y).
\ee

\subsection{Irreducible representations}

Let us denote by $C_\theta, \theta\in[0,\pi]$ the conjugacy classes in
$\SU(2)$, and by $g_\theta$ a representative of $C_\theta$ that is 
in the Cartan subgroup $U(1)$. The irreducible representations are
labeled by pairs $(\theta,k)$ of conjugacy classes and
representations of the centralizer $U(1)$. The carrier space $V^{(\theta,k)}$ is:
\be
V^{(\theta,s)} := \{ \phi:G\to \CC | \phi(x h) = e^{-i s\xi} \phi(x),
\forall x\in G, h={\rm diag}(e^{i\xi},e^{-i\xi}) \}
\ee
and the action of an element $f\in D(G)$ is:
\be
(\pi^{(\theta,s)}(f) \phi)(x) = \int dz\, f(x g_\theta
x^{-1},z)\phi(z^{-1} x).
\ee 
An orthonormal basis in $V^{(\theta,s)}$ is given by the matrix elements:
$\sqrt{{\rm dim}_j} T^j_{ms}(x), j\geq s, -j\leq m\leq j$.  We shall
also use the bra-ket notation and denote the basis vectors by
$|jm\rangle$. The carrier space $V^{(\theta,s)}$ can be decomposed  
into finite dimensional subspaces:
\be
V^{(\theta,s)} = \oplus_{j\geq s} V^{(\theta,s)}_j, \qquad
V^{(\theta,s)}_j = {\rm Span}\{ |jm\rangle \}.
\ee

\subsection{Matrix elements and characters}

Let us consider the matrix elements:
\be\label{mm-1}
(\pi^{(\theta,s)}(f))^{jj'}_{mm'} = \langle jm| \pi^{(\theta,s)}(f)
|j'm'\rangle = \\ \nonumber \sqrt{{\rm dim}_j {\rm dim}_{j'}} \int dx dz\,
\overline{T^j_{ms}(y)} f(y g_\theta y^{-1}, z) T^{j'}_{m's}(z^{-1}y).
\ee
By definition, the matrix elements are in $D(G)^*$. Using the pairing
$\langle\cdot,\cdot\rangle$ we can identify them with functions on
$G\times G$. To this end, let us transform the above formula. Let us
introduce, for each element $x\in C_\theta$ of the conjugacy class $C_\theta$, an 
element $B_x\in G$ such that:
\be
B_x g_\theta B_x^{-1}=x.
\ee
We will also need a $\delta$-function $\delta_\theta(x)$ picked on a conjugacy class $\theta$, which is
defined as:
\be\label{delta-t}
\int dg\, \delta_\theta(g) f(g) = \int_{G/H} f(x g_\theta x^{-1}) dx.
\ee
Using these objects the matrix element can be written as:
\be\label{mm-2}
(\pi^{(\theta,s)}(f))^{jj'}_{mm'} = \sqrt{{\rm dim}_j {\rm dim}_{j'}} \int dx dy\,
 f(x, y) \delta_\theta(x) \overline{T^j_{ms}(B_x)} T^{j'}_{m's}(y^{-1} B_x).
\ee
Therefore we define:
\be\label{me}
(\pi^{(\theta,s)})^{jj'}_{mm'}(x,y) := \sqrt{{\rm dim}_j {\rm dim}_{j'}} 
\delta_\theta(x) \overline{T^j_{ms}(B_x)} T^{j'}_{m's}(y^{-1} B_x).
\ee
This is the main formula for matrix elements as functions on $G\times G$.

The character is obtained in the usual way as the trace:
\be\label{char}
\chi^{(\theta,s)}(x,y) := \sum_{jm} (\pi^{(\theta,s)})^{jj}_{mm}(x,y) = 
\delta_\theta(x) \sum_j {\rm dim}_j T^{j}_{ss}(B_x^{-1} y^{-1} B_x).
\ee
To simplify this further let us use the fact that:
\be
\sum_j {\rm dim}_j T^j_{ss}(g) = e^{-is(\phi+\psi)}
\delta(\cos{\theta}).
\ee
We have used the Euler parameterization here:
$g=g(\phi,\theta,\psi)$. Applied to \eqref{char} this formula implies that 
$y=B_x g_\xi B_x^{-1}$, or, in other words, that \eqref{char} contains
the $\delta$-function $\delta_{xy}(yx)$. Thus, we have:
\be
\chi^{(\theta,s)}(x,y) = \delta_\theta(x) \delta_{xy}(yx)
\chi_s(B_x^{-1} y^{-1} B_x),
\ee
where $\chi_s(n)=e^{is\xi}, n={\rm diag}(e^{i\xi},e^{-i\xi})$. This is
the character formula given in \cite{Koorn}. 

\subsection{Decomposition of the identity}
Now that we have obtained the characters, we can use them to decompose
the identity element in the Drinfeld double. To this end, we will need
the Weyl integration formula:
\be
\int_G dg\, f(g) = \int_{H/W} \frac{d\theta}{\pi} \, \Delta(\theta)^2 \int_{G/H} dx\, f(x
g_\theta x^{-1}).
\ee
Together with \eqref{delta-t} this implies:
\be\label{ident-1}
 \int_{H/W} \frac{d\theta}{\pi} \, \Delta(\theta)^2 \delta_\theta(g) =
 1.
\ee
Let us now apply the summation over $s$ to \eqref{char}. We get:
\be\label{1-decomp}
 \int_{H/W} \frac{d\theta}{\pi} \, \Delta(\theta)^2 \sum_s \chi^{(\theta,s)}(x,y)=\delta_e(y).
\ee

\section{Group field theory for the Drinfeld double}
\label{sec:gft}

From the algebraic formulation of Boulatov's theory in section \ref{sec:Boul} it was clear that
the main object that is used in the construction of the model is the projector 
$\theta\in{\mathcal A}^{\otimes 3}$. Unlike the case of Boulatov theory,
where there is essentially a unique such projector constructed from
the identity element of $\mathcal A$, in the Drinfeld double case
there are several interesting ``identity elements'' that can be used
in the construction of $\theta$. We shall analyze several
different possibilities.

An explanation of why different possibilities can arise is as
follows. The $\theta$-projector that defines GFT should be constructed from a
projector on $\D$. Characters of irreducible representations are
projectors. To construct a more general projector one can take various linear combinations of
characters. Natural examples are given by: (i) the sum of
characters of all the representations of $\D$, which gives the
identity operator; (ii) the sum of characters of all simple
representations; (iii) the character of the trivial
representation. One can more generally consider not characters but
individual matrix elements, which are also projectors. In this paper we will 
consider the simplest case of the projector constructed from the
identity operator, as well as another in a certain sense dual projector constructed from the
element $\delta_e(x)$.

\subsection{Projector $\theta$ constructed from the identity $\delta_e(y)$}

Let us first consider a direct analog of Boulatov model. Thus, we
construct a projector $\theta$ from the identity element
$\delta_e(y)$ on $\mathcal D$.
To construct the projector we follow the same procedure that was used in section \ref{sec:Boul}. Namely,
let us first construct the propagator (or the kernel of the identity
operator $\delta_e(y)$):
\be
{\mathcal D}^* \otimes {\mathcal D}^* \ni G = (\kappa\otimes{\rm id})\Delta^* 1, \\ \nonumber
G(x_1,y_1;x_2,y_2) = \delta_e(x_1 x_2) \delta_e(y_1 y_2^{-1}).
\ee
Let us define an operator 
$h_3:{\mathcal D}^* \otimes {\mathcal D}^* \otimes {\mathcal D}^* \to \CC$ via:
\be\label{h3}
&{}& h_3(f_1,f_2,f_3):=h^*(f_1\star f_2\star f_3) = \\ \nonumber
&{}&\int dz_1 dz_2 dz_3 dy \,\, f_1(z_1,y)
f_2(z_2,y) f_3(z_3,y) \delta_e(z_1 z_2 z_3).
\ee
The $\theta$-projector is obtained by applying $h_3$ to $G^{\otimes 3}$:
\be
\theta:=({\rm id}^{\otimes 3}\otimes h_3)\, G\otimes G\otimes G.
\ee
Let us find an explicit expression for this projector. We have:
\be\label{theta*}
\theta(x_1,y_1;x_2,y_2;x_3,y_3) =  
\int dz_1 dz_2 dz_3 dy \,\, \delta_e(x_1 z_1) \delta_e(x_2 z_2) \delta_e(x_3 z_3) \times \\ \nonumber
\delta_e(z_1 z_2 z_3) \delta_e(y_1^{-1} y)\delta_e(y_2^{-1} y)\delta_e(y_3^{-1} y)= 
\delta_e(x_1^{-1} x_2^{-1} x_3^{-1}) \theta(y_1,y_2,y_3),
\ee
where $\theta(y_1,y_2,y_3)$ is given by \eqref{theta}. An explicit verification shows that
$\theta$ is real $\theta^*=\theta$. We also have to check the
projector property of $\theta$. However, unlike the case of Boulatov
model, the quantity \eqref{theta*} is not a projector. The
$\bullet$-product of two $\theta$ gives a divergent factor of
$\eta^{-2}$ \eqref{eta} that we have already encountered before in the discussion
of the Ponzano-Regge model. We nevertheless proceed formally, and
introduce a multiple of $\theta$ as a projector. The
following lemma is a statement to this effect.
\begin{lemma}The quantity $\eta^2\theta$, where
  $\eta^{-2}=\delta_e(e)$ is a projector:
\be
(\eta^2 \theta)\bullet (\eta^2\theta) = \eta^2 \theta.
\ee
\begin{proof}
A proof is by verification. We have:
\be
(\theta\bullet\theta)(x_i,y_i)=\delta_e(x_1^{-1}x_2^{-1}x_3^{-1}) \\
\nonumber \int
dg\, \prod_i \int dz_i\,\delta_g(z_i) \int d\tilde{g}\, \delta_e(z_1^{-1}
x_1^{-1} z_1 z_2^{-1} x_2^{-1} z_2 z_1^{-1} x_1^{-1} z_1) \prod_i
\delta_{\tilde{g}}(z_i^{-1} y_i) = \\ \nonumber
(\delta_e(x_1^{-1} x_2^{-1} x_3^{-1}))^2 \int dg d\tilde{g}\, \prod_i
\delta_{\tilde{g}}(g^{-1} y_i) = \eta^{-2} \theta(x_i,y_i).
\ee
\end{proof}
\end{lemma}

The ``projector'' so defined is the weakest point of our
construction. A sceptic may argue that this quantity does not exist or
is zero. We note, however, that both such a projector and the $\eta$
factor would be perfectly
well-defined had we been working with the Drinfeld double of the
quantum $\SU_q(2)$ at root of unity instead. Passing to the quantum
$\SU_q(2)$ is known to have the interpretation of adding a
(positive) cosmological constant. Thus, as we see, unlike the Boulatov
model itself, the model for the Drinfeld double of $\SU(2)$ is defined
only formally. Later we shall consider another model which is
well-defined.

The quantity $\theta$ can be expressed in terms of the $R$-matrix. To state a result to this effect, we
define the following graphical notation:
\be
R(\phi\otimes\psi)(x_1,y_1;x_2,y_2)=\phi(x_1,y_1)\psi(x_1^{-1}x_2 x_1,x_1^{-1} y_2) =\,\,
\lower0.4in\hbox{\epsfig{figure=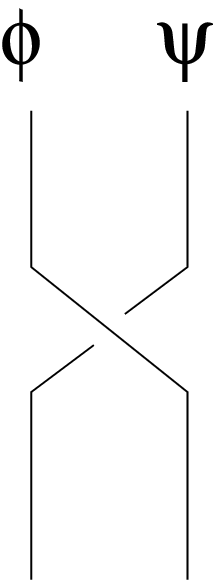, height=0.8in}}
\ee
The opposite braiding is given by:
\be
\sigma R(\phi\otimes\psi)(x_1,y_1;x_2,y_2)=\phi(x_2^{-1} x_1 x_2, x_2^{-1} y_1)\psi(x_2,y_2) =\,\,
\lower0.4in\hbox{\epsfig{figure=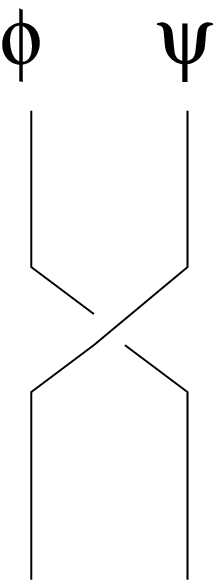, height=0.8in}}
\ee
The following result is confirmed by an explicit computation:
\begin{lemma}
\be\label{r-lemma}
\lower0.3in\hbox{\epsfig{figure=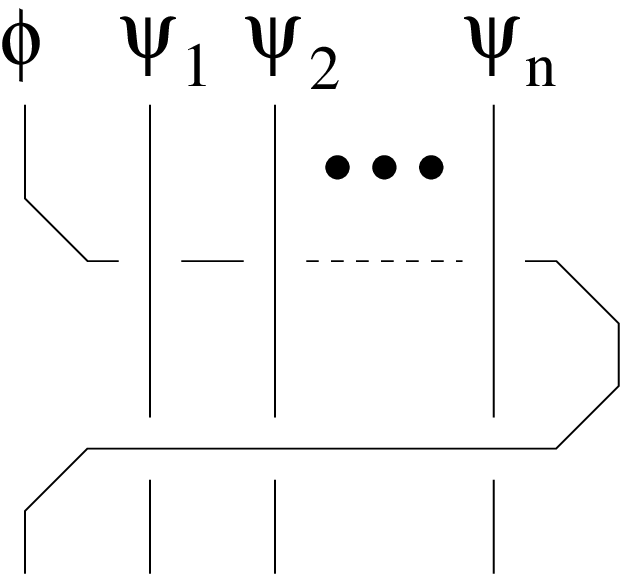, height=0.8in}} \,\,= 
\phi(x^{-1} x_1^{-1}\ldots x_n^{-1}x x_n \ldots x_1 x, 
x^{-1} x_1^{-1} \ldots x_n^{-1}xy) \times \\ \nonumber
\prod_{i=1}^n \psi_i(x^{-1} x_i x, x^{-1} y_i).
\ee 
\end{lemma}
Let us now take $\phi(x,y)=\delta_e(y)$, and take the Haar functional 
$h$ in the first channel. We get the following result:
\begin{lemma}
\be\label{general-braid}
\lower0.3in\hbox{\epsfig{figure=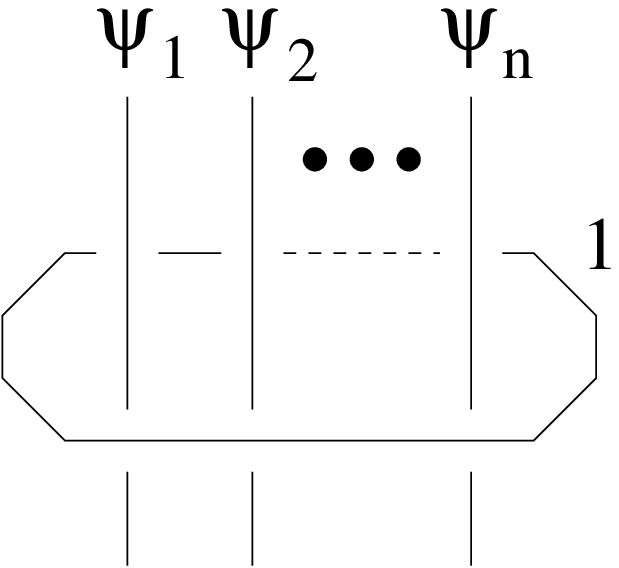, height=0.8in}} \,\,= 
\delta_e(x_1^{-1}\ldots x_n^{-1}) \int dg \prod_i \psi_i(g^{-1} x_i
g,g^{-1} y_i).
\ee
\end{lemma}
We can use this result to give another description of the quantity $\theta$:
\begin{lemma} \label{lemma-1}
\be
\lower0.25in\hbox{\epsfig{figure=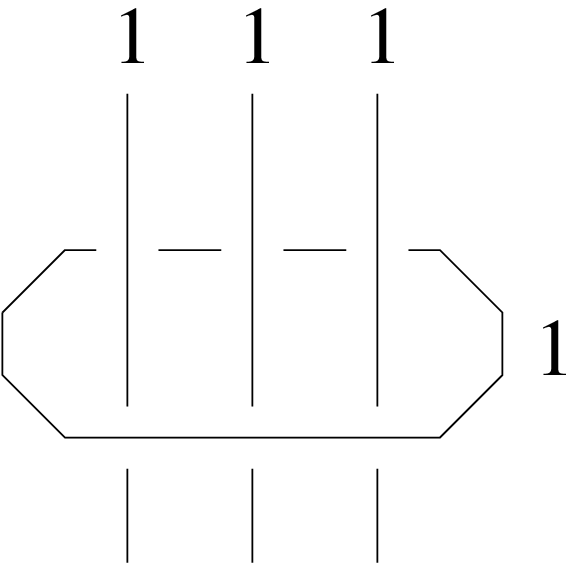, height=0.8in}} \,\,= \theta
\ee
\end{lemma}

Let us consider the object \eqref{general-braid}.
We will need two properties of such objects, which are
referred to as the {\it handleslide} and {\it killing}:
\begin{lemma} A composition of two quantities \eqref{general-braid}
 satisfies the following ``handleslide'' property:
\be\label{handleslide}\label{lemma-2}
\lower0.4in\hbox{\epsfig{figure=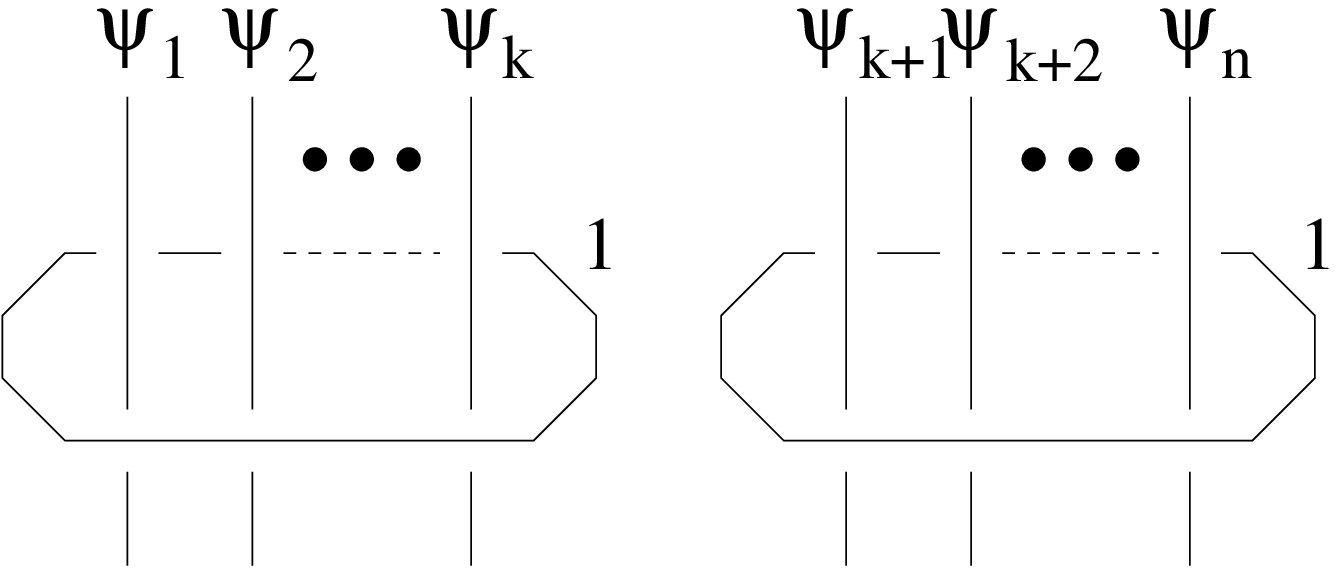, height=0.8in}}=
\lower0.55in\hbox{\epsfig{figure=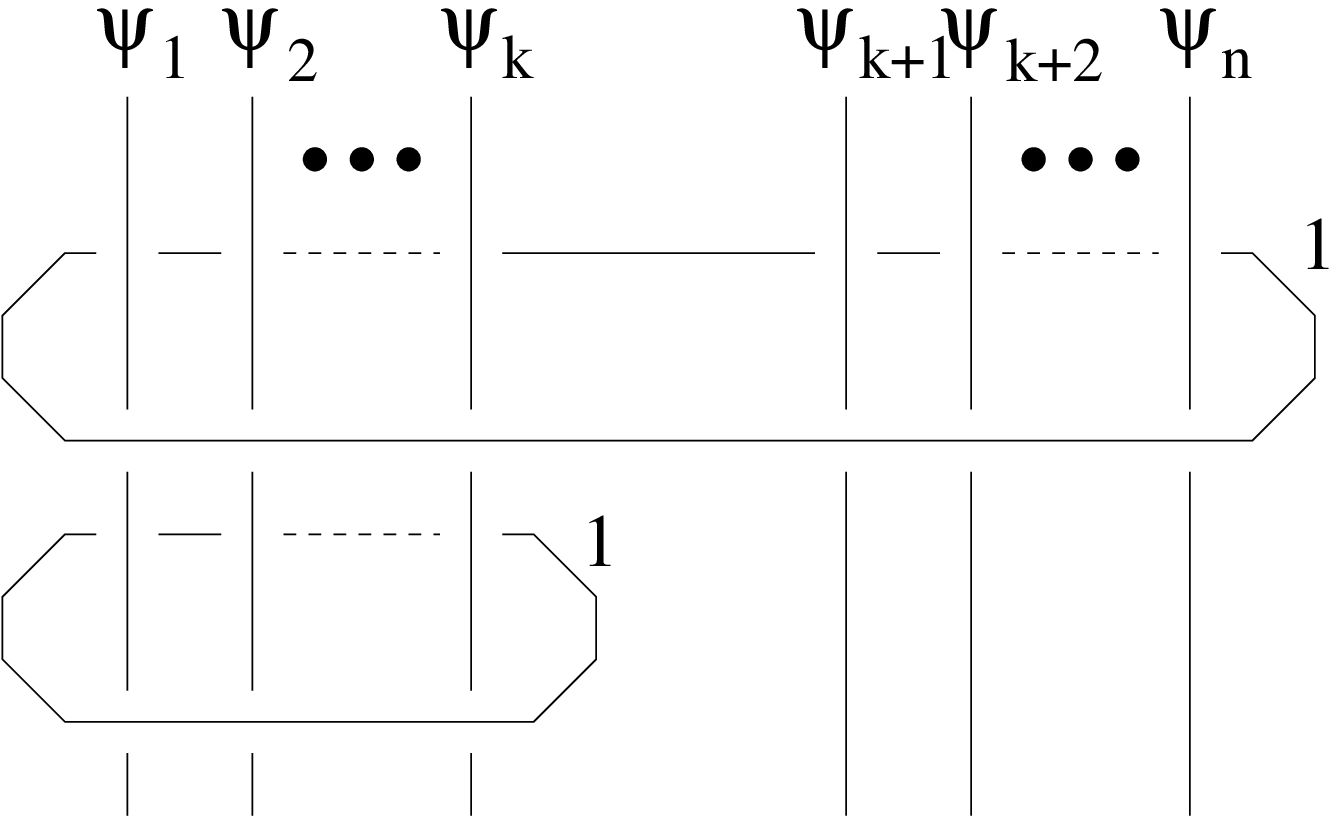, height=1.1in}}
\ee
\begin{proof}A direct computation of the left hand side gives:
\be\label{handle-1}
\delta_e(x_1^{-1} \ldots x_k^{-1}) \int dg \, \prod_{i=1}^k
\psi_i(g^{-1} x_i g,g^{-1} y_i) \times \\ \nonumber
\delta_e(x_{k+1}^{-1} \ldots x_n^{-1}) \int d\tilde{g} \, \prod_{i=k+1}^n
\psi_i(\tilde{g}^{-1} x_i \tilde{g},\tilde{g}^{-1} y_i).
\ee
The right-hand side is given by:
\be\label{handle-2}
\delta_e(x_1^{-1}\ldots x_n^{-1}) \delta_e(x_1^{-1}\ldots x_k^{-1}) \\
\nonumber 
\int dgd\tilde{g} \, \prod_{i=1}^k \psi_i(g^{-1} \tilde{g}^{-1} x_i
\tilde{g} g, g^{-1} \tilde{g}^{-1} y_i) \prod_{i=k+1}^n \psi_i(g^{-1}
x_i g, g^{-1} y_i).
\ee
A straightforward change of integration variable in the first part of
the product in \eqref{handle-2} makes it identical to \eqref{handle-1}
and proves the lemma.
\end{proof}
\end{lemma}

\begin{lemma} \label{lemma-3} The following killing property holds:
\be\label{killing}
\lower0.3in\hbox{\epsfig{figure=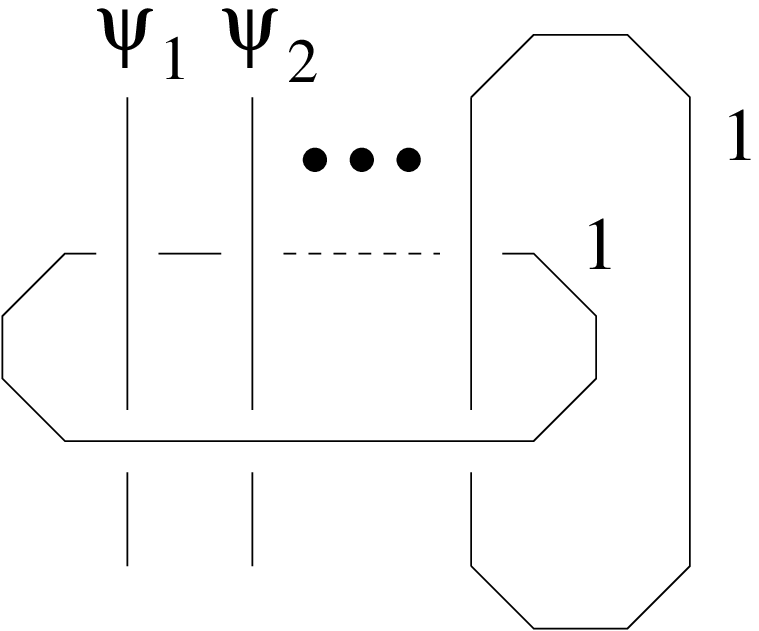, height=0.8in}}=
\lower0.3in\hbox{\epsfig{figure=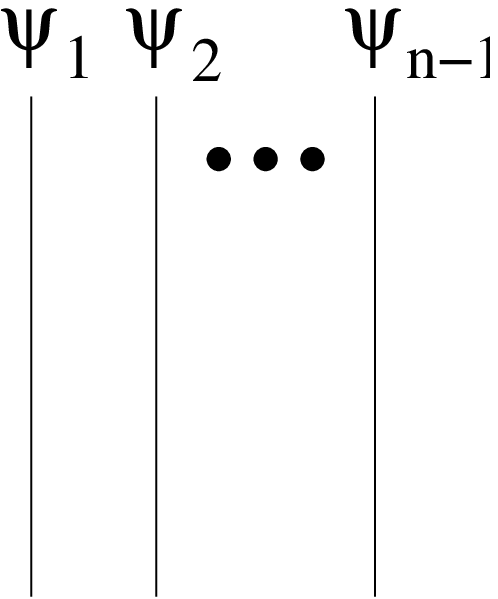, height=0.8in}}
\ee
\begin{proof} We take $\psi_n(x_n,y_n)=\delta_e(y_n)$, 
  compute \eqref{general-braid}, and apply the Haar functional $h$ in
  the last channel. We get:
\be
\int dx_n \delta_e(x_1^{-1} \ldots x_n^{-1}) \int dg \,
  \prod_{i=1}^{n-1} \psi_i(g^{-1} x_i g,g^{-1}y_i) \delta_e(g^{-1})=
\prod_{i=1}^{n-1} \psi_i(x_i,y_i).
\ee
This proves the lemma.
\end{proof}
\end{lemma}

For completeness, let us also give a decomposition of the quantity $\theta$ into
characters:
\begin{lemma} The following decomposition of the quantity $\theta$
  holds:
\be
&{}&\theta(x_1,y_1;x_2,y_2;x_3,y_3)= \int_{H/W} \prod_{i=1}^3 \frac{d\theta_i}{\pi} \, \Delta(\theta_i)^2 \times
\\ \nonumber
&{}& \sum_{j_1,j_2,j_3} {\rm dim}_{j_1} {\rm dim}_{j_2} {\rm dim}_{j_3} 
\theta^{\theta_1,\theta_2,\theta_3}_{j_1,j_2,j_3}(x_1,y_1;x_2,y_2;x_3,y_3),
\ee
where
\be\label{theta-char}
&{}&\theta^{\theta_1,\theta_2,\theta_3}_{j_1,j_2,j_3}(x_1,y_1;x_2,y_2;x_3,y_3)= \\ \nonumber
&{}&\delta_{\theta_1}(x_1) \delta_{\theta_2}(x_2) \delta_{\theta_3}(x_3) 
\delta_e(x_1^{-1} x_2^{-1} x_3^{-1}) \theta_{j_1,j_2,j_3}(y_1,y_2,y_3), 
\ee
where $\theta_{j_i}(y_i)$ is given by \eqref{theta-j}.
\begin{proof}
Let us use the character decomposition of the identity operator $\delta_e(y)$:
\be
\delta_e(y) = \int_{H/W} \frac{d\theta}{\pi} \Delta(\theta)^2 \sum_j
      {\rm dim}_j \,\chi^\theta_j(x,y),
\ee
where
\be\label{char*}
\chi^{\theta}_j(x,y) := \sum_{s,m} (\pi^{(\theta,s)})^{jj}_{mm}(x,y) = 
\delta_\theta(x) \chi_j(y).
\ee
We can now apply the formula \eqref{general-braid} to a product of
3 characters, which results in \eqref{theta-char}.
\end{proof}
\end{lemma}

\subsection{Projector constructed from $\delta_e(x)$}

Another projector that we consider in this paper will be built from
the operator $\hat{1}(x,y)=\delta_e(x) \in\D$. This element is
{\it not} an identity: $(\hat{1} \bullet f)(x,y)= \delta_e(x) h^*(f)$.
As we see, it plays the role of the element dual to the Haar measure
on $\D^*$. We have put a hat above the symbol
denoting this operator in order not to confuse it with the identity
$\delta_e(y)$ in the Drinfeld double. It is clear that the
$\theta$-projector constructed from $\hat{1}$ according to the rules
as described above is equal to $\eta^{-2} \hat{1}^{\otimes 3}$. Indeed, we take
3 operators $\hat{1}$ and enclose them with a strand with $\hat{1}$
inserted. The result \eqref{r-lemma} proves the assertion. It is
obvious that a multiple of $\theta$ constructed this way is (formally) a
projector. Another important property of the operator $\hat{1}$ is that
it satisfies the handleslide and killing properties. Let us state two lemmas to this effect.
\begin{lemma} \label{lemma-h} When the projector $P=\hat{1}$ is inserted
  into the meridian link, the handleslide property holds.
\begin{proof}
The left hand side is given by:
\be
\delta_e( x_1^{-1} \ldots x_k^{-1}) \prod_{i=1}^k \psi_i(x_i, y_i)
\delta_e( x_{k+1}^{-1} \ldots x_n^{-1}) \prod_{i=k+1}^{n} \psi_i(x_i,y_i).
\ee
The right-hand side is given by:
\be
\delta_e( x_1^{-1} \ldots x_k^{-1} x_{k+1}^{-1} \ldots x_n^{-1}) \prod_{i=1}^k \psi_i(x_i, y_i)
\delta_e( x_{k+1}^{-1} \ldots x_n^{-1}) \prod_{i=k+1}^{n} \psi_i(x_i,y_i).
\ee
The handleslide property is obvious.
\end{proof}
\end{lemma}

\begin{lemma} \label{killing-h} The quantity $\hat{1}$ satisfies the
  killing property provided the operator inserted into the longitude
  is a function of $y$ only.
\begin{proof} A proof is by direct verification.
\end{proof}
\end{lemma}

Having discussed several different projectors that can be used in the
construction we are ready to define the model.

\subsection{The model}

There are now several possible models that can be formulated, depending on
which $\theta$-projector one uses. We shall proceed at the formal
level, introducing factors of $\eta$ when necessary. Later the models
we obtain will be placed on a more solid footing using the chain mail
techniques. 

Let us consider a projector operator $\eta^2 \theta$.
The basic field of the model
is $\phi\in {\mathcal D}^{\otimes 3}$ that is real $\phi^*=\phi$, and
one builds a  projected field $\tilde{\phi}$ via:
\be\label{theta-phi}
&{}&\tilde{\phi}:=\eta^2 \theta\bullet \phi=\eta^2 \,\,
\lower0.25in\hbox{\epsfig{figure=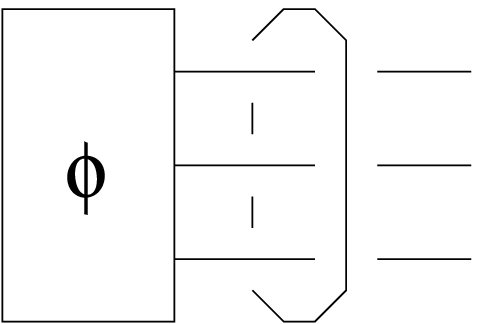, height=0.5in}}
\ee
In this graphical notation what is inserted on the ``meridian'' link
and other 3 strands depends on the model. In all cases $\eta^2 \theta$
is a projector. Using the graphical notation introduced, the
action for the model is defined as follows:
\be\label{action*}
S[\phi] = \frac{\eta^2}{2} \,\,\, \lower0.25in\hbox{\epsfig{figure=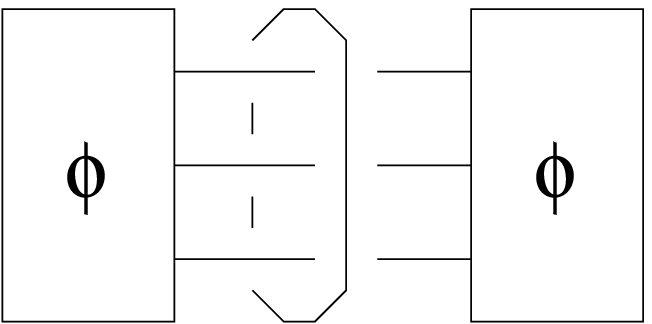, height=0.5in}}
+ \frac{g}{4!}  \eta^8 \,\,\, \lower1in\hbox{\epsfig{figure=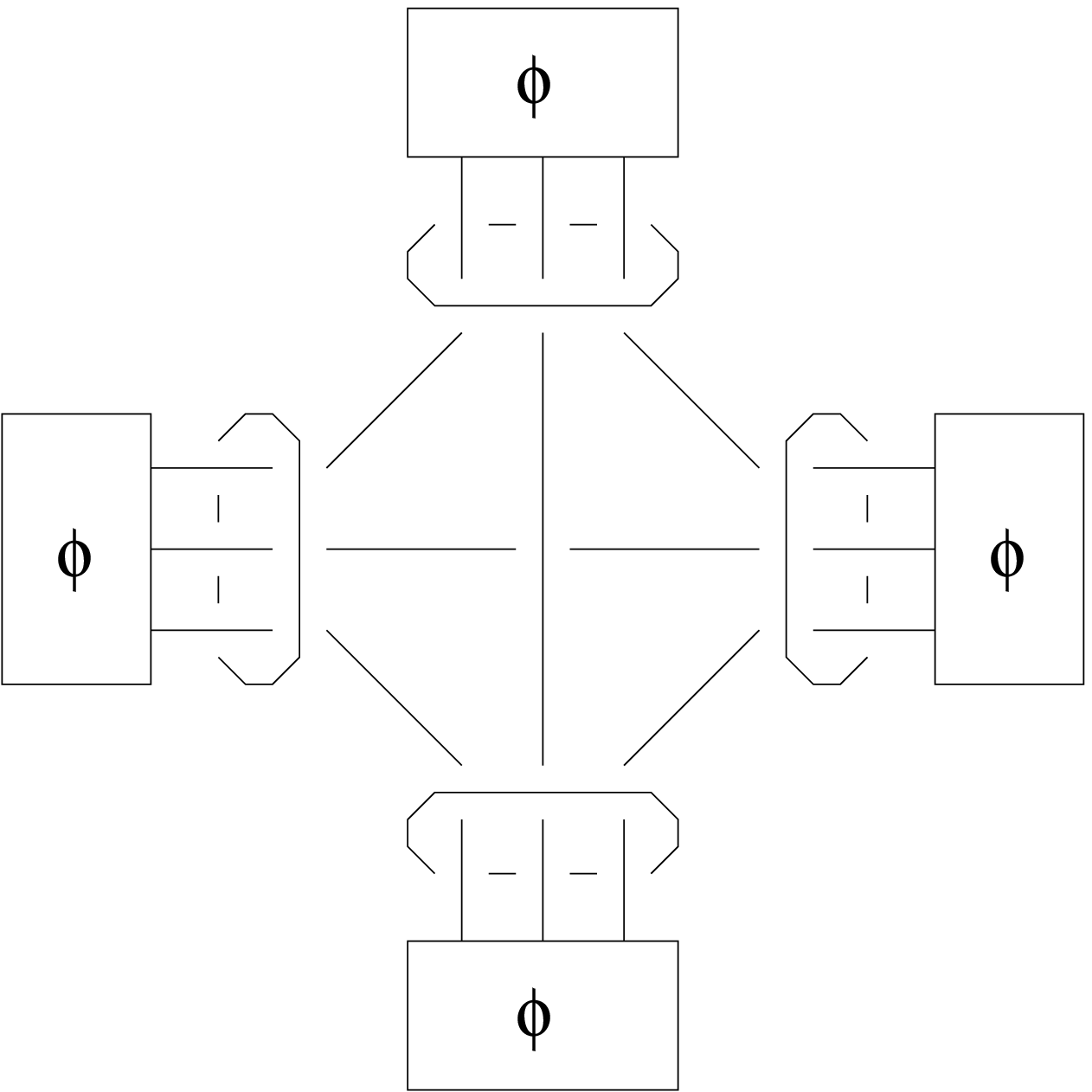, height=2in}}
\ee

To describe Feynman amplitudes generated by these models we shall introduce a notion of
the chain mail.

\subsection{Roberts' chain mail}

Using an idea of chain mail Roberts presented \cite{Roberts} a very convenient description of the
Turaev-Viro invariant. Here we remind the reader this notion, and show that the Feynman
amplitude generated by our model is just a chain mail evaluation.

Let us consider the Feynman diagram perturbative expansion of the
model \eqref{action*}. As the propagator of the model consists of
3 lines, each diagram is a collection of vertices (0-cells), edges
(1-cells) and faces (2-cells) obtained by following each line
till it closes. Each diagram $D$ is an abstract one, that is not embedded
in any space. Note that the data of a diagram $D$ define a
handlebody $H(D)$, which is just the blow up of the graph of $D$.
In other words, to obtain $H(D)$ one takes solid balls (0-handles), one for every
vertex of the diagram, and attaches them to each other by solid
cylinders (1-handles), one cylinder for every edge. The handlebody $H(D)$ is not
embedded in any space. 

Now draw the curves $\epsilon_i$ defining the 2-faces on the
boundary $\partial H(D)$, and push them slightly inside of $H(D)$. Let us
then add the meridian curves $\delta_i$ for all the 1-cells. The
obtained collection of curves is called a {\it chain mail link}
$C(D)\subset H(D)$ of $D$. We now attach an appropriate
(model-dependent) operator from $\mathcal D$ to every component of the
link $C(D)$, and evaluate the link to obtain a value $\Omega C(D) \in
\CC$. This evaluation procedure is as follows. One views the
chain mail as a rule for taking a product of $\theta$-projectors, one
$\theta$ for every 1-cell. The projectors are $\bullet$-multiplied as specified
by the 0-cells. Every closed loop corresponds to applying the Haar
functional $h$. This evaluation rule will become more clear when we consider concrete examples
below. We note that this evaluation is {\it not} an evaluation
of the chain mail as embedded in some 3-manifold. It can be related to
the more standard evaluation of the chain mail as embedded (as appears
in Roberts' work \cite{Roberts}), see more remarks on this below. At
the moment, however, there is no 3-manifold associated to any Feynman diagram.

\begin{proposition} \label{prop} For the model \eqref{action*} the amplitude
  $\mathcal Z$ of a  Feynman diagram $D$ is equal to $g^V \eta^{2E}
  \Omega C(D)$, where $V$ is the number of vertices and $E$ is the
  number of edges of $D$. Feynman amplitude of the model constructed
  using $\hat{1}$ is equal to $g^V \Omega C(D)$.
\begin{proof} In the space of projected fields
  $\tilde{\phi}=\theta\bullet\phi$ the kinetic operator $\eta^2
  \theta$ is the identity operator. Its inverse is itself. Thus, the
  propagator of the model  \eqref{action*} is
  again $\eta^2 \theta$, and the vertex is $g$ times the operator
  depicted graphically in \eqref{action*}. This set of Feynman rules
  makes the first statement evident. The second case is similar,
  except that no factors of $\eta$ appear.
\end{proof}
\end{proposition}

\subsection{Feynman diagrams and 3-manifolds}

\begin{figure}\label{fig:trun}
\begin{center}
\epsfig{figure=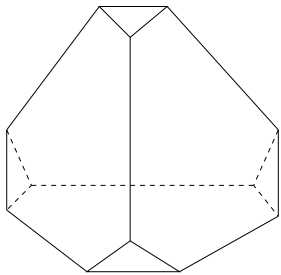, height=1.4in}
\end{center}
\caption{A truncated tetrahedron}
\end{figure}

Each Feynman diagram $D$ defines a 3D pseudo-manifold $PM(D)$. It is easiest to obtain
$PM(D)$ by gluing together truncated tetrahedra an example of which is shown in Fig. \ref{fig:trun}.
Thus, each vertex of $D$ naturally corresponds to a truncated tetrahedron. Each edge of $D$ defines a 
gluing of two truncated tetrahedra $T,T'$: a large face of $T$ is glued to a large face of $T'$. This way 
each (closed) Feynman diagram $D$ defines a 3D pseudo-manifold $PM(D)$ with boundary $\partial PM(D)$. The
pseudo-manifold $PM(D)$ can be completed to a manifold $M(D)$ if the boundary $\partial PM(D)$ is a disjoint
union of a number of spheres $S^2$. In this case $M(D)$ is formed by gluing to $PM(D)$ a number of 3-balls. 
See \cite{Roberto} for more details on this construction.

Note, however that it may well happen for some of the Feynman diagrams that $\partial PM(D)$ contains
boundary components other than spheres. In this case one needs extra structure to be added to the
Feynman diagram to ``convert'' it into a 3-manifold. This extra structure is easiest to understand
on the example of a toroidal boundary component in $\partial PM(D)$. In this case, to convert the
3-manifold $PM(D)$ with boundary into a closed 3-manifold one has to glue in a solid torus. However,
there is no unique solid torus. A solid torus is specified by a closed curve on its boundary (torus) which
is contractible inside the solid torus. Such a curve is in turn specified by two relatively prime
numbers. This is exactly the extra data (for each toroidal boundary component) 
that are necessary to convert a Feynman diagram into 
a closed 3-manifold in cases when the manifold obtained by gluing in the two handles contains
toroidal boundary components. When the boundary components are of higher genus, one needs
even more data, which are a set of $3g-3$ curves on the boundary ($g$ is the genus of the boundary component
in question) that are contractible inside the handlebody one glues in. These issues are delicate ones,
and we would not like to divert the reader from the main theme by pursuing them any further. 
Let us just say that one either chooses (if possible) the symmetry properties of the model so that 
the boundary of $PM(D)$ is always a collection of spheres, or, if the former is not possible, 
adds some extra structure to the model that tells one how to complete the Feynman diagrams
into 3-manifolds. We will assume that one of the two is done, and proceed without worrying
about these issues any more.

Let us now explain a relation to the chain mail construction of the previous subsection is as follows. One obtains $PM(D)$ 
by taking the handlebody $H(D)$ and gluing to it a number of 2-handles (solid cylinders). The
gluing is done as specified by the $\epsilon_i$ curves. Once $PM(D)$
is obtained, one glues a number of 3-handles to obtain $M(D)$. After this,
using the ``extra rules'' discussed in the previous paragraph, one
glues in a collection of 2-spheres or higher genus handlebodies, and completes $PM(D)$
to a 3-manifold $M(D)$ without boundary. The original Feynman diagram $D$ is
now embedded into $M(D)$. The evaluation that we described above did
not need a representation of $D$ as embedded into some 3-manifold, and
in that sense was independent of any embedding. However, once
the diagram is embedded into $M(D)$, the evaluation that determines
the Feynman amplitude can also be thought as the usual Reshetikhin-Turaev-Witten
evaluation of $D$ in $M(D)$. The result of this evaluation does depend on
the embedding of $D$ in $M(D)$. The amplitude one gets is thus that
for a process described by the diagram $D$, happening in a particular
3-manifold $M(D)$. This should be contrasted with the usual situation
in QFT, where the Feynman diagrams are not embedded in any space. The
extra structure of faces and ordering of the faces that naturally follows
from the group field theory expansion is exactly the structure that
allows to think of $D$ as embedded in a 3-manifold. The 3-manifold
that gets constructed from $D$ with its extra structure is what
describes the gravitational part of the degrees of freedom of the model.

We now note that different Feynman graphs can lead to one and the same topological
manifold $M(D)$. There are certain moves that relate different $D$
that correspond to the same manifold $M$. As stated by the following
theorem, when the operator that is inserted on the strands satisfies
the handleslide and killing properties, the chain mail evaluation is
invariant under the moves and thus gives an invariant of $M$.

\begin{theorem} (Roberts) Assume that the operator ${\mathcal O}$ inserted on links
  forming  $C(D)$ satisfies the handleslide and killing properties. Assume
  that the trace of this operator is given by $\eta^{-2}$. The
  following multiple of the chain mail evaluation:
\be\label{ch}
CH(D):=\eta^{2n_0 + 2n_3} \Omega C(D),
\ee
  where $n_0$ and $n_3$ are the numbers of 0 and 3-handles correspondingly,
  is an invariant of 3-manifolds in that $CH(D)=CH(D')$ when $M(D)\sim
  M(D')$ are 3-manifolds of the same topology.
\begin{proof} We shall only present a sketch, as a detailed proof is
  given in \cite{Roberts}. Different Feynman diagrams that correspond
  to the same $M$ are different handle decompositions of $M$. A known
  theorem of topology states that different handle decompositions can
  be related by a sequence of births of $k, k-1$-handle pairs, and by
  handleslides. Invariance under handleslides is guaranteed by the the
  handleslide property of the operator inserted along the strands. To
  analyze the birth of 1-0 and 3-2 handle pairs one uses the
  handleslide property, and is left with a single unknot with
  ${\mathcal O}$ inserted and not linked with the rest of the chain
  mail. This gives a factor of $\eta^{-2}$ absorbed by the prefactor
  of $CH(D)$.  A birth of the 2-1 handle pair is handled using the
  killing property.
\end{proof}
\end{theorem}

\begin{remark} Note that if the operator inserted along the meridians
  and longitudes satisfies the handleslide and killing properties, and
  its trace is equal to unity, then a similar theorem holds, except
  that it is now the chain mail evaluation itself that is invariant.
\end{remark}

\begin{corollary}The model constructed using
the identity operator $\delta_e(y)$ gives invariants of
3-manifolds. 
\end{corollary}

\begin{remark} There is another model that gives manifold invariants, namely the model 
 constructed using $\hat{1}$. This model, however, requires a somewhat different set of
formal prefactors. We shall describe it in details below. 
\end{remark}

\section{More general models}
\label{pp-models}

In the previous section we have seen how group field theory for the
Drinfeld double leads to the notion of chain mail in that Feynman diagrams of GFT
are computed as the chain mail evaluation. Chain mail arises because
operators that are used in the construction of the GFT model are
projectors. One can therefore multiply the $\theta$-projectors
appropriately, and be left with a chain mail where there is just one
meridian link per edge of a Feynman diagram $D$. Also importantly, the
operator that is inserted in the strands is a projector, and after
strands close combines to give a single operator in the longitude. 

One can consider a more general class of models not related to any
GFT, but formulated directly in terms of the chain mail. Thus, one
takes a chain mail that corresponds to a graph $D$, inserts one
species of operators into the meridians, some other species into
the longitudes, and evaluates the resulting link. The operators
inserted do not have to be projectors anymore, and this is what makes such
models different from the GFT ones. However, as we shall see, these more
general models admit a physical interpretation and are of interest. We
could have directly started from the notion of chain mail and
formulated all models correspondingly. However, we believe that the
GFT description we have given is essential in that it clear shows what
is and what is not possible in the GFT framework. The GFT models
described are also of importance in view of possible generalization to
algebras other than $\D$. Thus, for instance, the analog of Boulatov
model for the quantum group $\SU_q(2)$ has not been formulated. 
It is straightforward to do so using our algebraic formulation
described above.

\subsection{Formulation of the models}

Let us define a set of models as follows.
Consider a chain mail evaluation in which an operator $P=\delta_e(x) P(y)$ is
inserted in all the longitudes. The function $P(y)$ that appears as part
of this operator is not required to be a projector. Instead, we will just require $P(y)$ to be
normalized so that $\int dy P(y)=1$. Let us consider the following model.
\begin{definition} The amplitude of the model is
  obtained by inserting the identity $\delta_e(y)$ into the meridian
  links, and $\delta_e(x) P(y)$ into the longitudes. 
The amplitude is a (formal) multiple of the chain mail evaluation: 
\be\label{gen-model}
CH'(D)=\eta^{2n_1} \Omega C(D).
\ee
Here $\Omega$ denotes the evaluation, and the factor of $\eta^2$ to the power of the 
number of 1-handles is introduced for reasons that will become clear below.
\end{definition}

Let us prove some properties of the objects that the both models are 
constructed from. First let us consider a
number of strands enclosed by the $P$ operator. Such an object is
given by:
\be\label{p-1}
P(x_1^{-1} \ldots x_n^{-1}) \prod_i \psi_i(x_i, y_i).
\ee
Importantly, there is no longer a handleslide property \ref{lemma-2} for 
objects \eqref{p-1}, as is easy to verify. However, the killing property still
holds. Indeed, we take $\psi_n=\delta_e(y)$, and apply the Haar functional
in the last channel. We get:
\be
\int dx_n P(x_1^{-1} \ldots x_n^{-1}) \prod_{i=1}^{n-1} \psi_i(
x_i, y_i) \delta_e(e) = \\ \nonumber \eta^{-2} \int dg P(g) \, \prod_{i=1}^{n-1} \psi_i(x_i,y_i) = 
\eta^{-2} \, \prod_{i=1}^{n-1} \psi_i(x_i,y_i) ,
\ee
where we made a change of integration variable to arrive to the second expression. 

Using these facts, it is easy to prove the following assertion.
\begin{proposition}\label{prop*} Amplitudes given by \eqref{gen-model} are invariant
  under 0-1 and 2-1 handle pair births/deaths, as well as under 1-handle slides.
\begin{proof} Proof is same as that of Roberts' theorem of the previous
  section. The factor of $\eta^{2n_1}$ in \eqref{gen-model} is 
  necessary to guarantee the 0-1 handle pair births/deaths
  invariance as well as the invariance under the 2-1 handle pair births/deaths. Note that in 
the case considered by Roberts, see (\ref{ch}) above, one needs two prefactors to guarantee
the 0-1 and 2-3 moves. In our case we do not require the 2-3 moves, but have an additional
factor of $\eta^{-2}$ appearing in the 1-2 moves. This makes the correct prefactor
to be $\eta^2$ to the power of $n_1$ not $n_0$ as in case of (\ref{ch}). 
\end{proof}
\end{proposition}

\subsection{A model giving 3-manifold invariants}

Let us note that the case $P(y)=1$ is special, because in this case there
is the 2-handleslide property. In this case the chain mail evaluation gives 3-manifold
invariants. We have the following (almost trivial) corollary.
\begin{corollary} Evaluate the chain mail by inserting the identity
  operator $\delta_e(y)$ in all meridian links, and the operator $\delta_e(x)$
  into all longitudes. The quantity \eqref{gen-model} is a topological 
invariant in that it only depends on the topology of $M(D)$. Moreover,
its value is one for any manifold $M(D)$.
\end{corollary}

\subsection{Interpretation of 2-handleslide invariance absence}

The example of the previous subsection shows that there are models
that are interesting but not in the class of GFT models considered in
the previous section. It is not hard to give a more flexible
definition of GFT that would cover more examples. We shall not attempt
this however, concentrating instead on the chain mail definition from
now on.  

General models defined via chain mail do not produce
3-manifold invariants because there is no handleslide property for the
operator inserted in the longitudes. Still, as we shall presently see,
from the point of view of point particle theory these models are quite
natural. To see this, let us analyze what
happens if there is no handleslide property. Meridian links correspond to 1-handles,
and the longitudes correspond to 2-handles. There is
the handleslide property for the identity operator, see lemma
\ref{lemma-2}. Thus, the chain mail evaluation $\Omega CH(D)$ is
invariant under the 1-0 handle pair births and deaths. The killing
property guarantees an invariance under the 2-1 pair births and
deaths. There is also invariance under the handleslides along
1-handles. Thus, non-invariance comes only from two sources: there is
no invariance under 2-handle slides, and there is no 3-2 handle pair
birth-death invariance. Now recall that each $D$ is the dual skeleton
of some triangulated 3-manifold $M(D)$. As we have discussed in the
introduction, we would like to interpret the triangulation as a
Feynman diagram as well. Edges of this Feynman diagram are in
one-to-one correspondence with 2-handles, and vertices with 3-handles
of $M(D)$. Absence of invariance under moves involving 2-handles just
means that the amplitude we get is not invariant under changes in the
Feynman diagram. The absence of invariance under 3-2 pair birth-death
means that the Feynman amplitude is not invariant under a vertex being
replaced by two vertices connected by a new edge. Similarly, the
absence of the 2-handleslide property means that one cannot move edges
through vertices along other edges. All these moves change the Feynman diagram, and
the fact that there is no invariance is very natural from the point of
view of point particle field theory. Thus, in this case the model
leads not to 3-manifold invariants, but to Feynman amplitudes of the
dual point particle field theory. To get most general point particle
models we will have to work with theories of this class.

\section{Evaluation and interpretation of the models}
\label{sec:pp}

We have considered a set of models. Some of them were shown to give
invariants of 3-manifolds, some other only give amplitudes for
dual Feynman graphs. To analyze the structure of the amplitudes
arising, and to give it a geometrical and
point particle interpretation, let us use the kernel representation. 

\subsection{Kernel representation}

Recall that in the kernel representation one
associates an operator $f(x,y)\in {\mathcal D}$
its kernel defined as $(\kappa\otimes{\rm id})\Delta^* f$. A simple computation gives:
\be\label{k-1}
f(x,y|\tilde{x},\tilde{y})=f(y^{-1} x^{-1} y, y^{-1} \tilde{y}) \delta_e(x \tilde{x}).
\ee
The kernels should be multiplied using the $\star$-product and the $h^*$ Haar functional should be taken. 
In the case of algebra of functions on the group this operation is just a point-wise multiplication
of the propagators with an integral taken. In the case of the Drinfeld double the structure is more involved.
To display it, we note that:
\be
h^*(f_1\star f_2) = \int dxdy\,\, f_1(x,y)f_2(x^{-1},y).
\ee
Thus, in order to compose kernels of two operators, one should take one kernel and multiply it
by the other kernel with $x$ replaced by $x^{-1}$, and then integrate over $x,y$. Equivalently, one
can replace $x$ by $x^{-1}$ in \eqref{k-1}. It is this quantity that
we shall refer to as the operator kernel:
\be\label{k-2}
K_f(x,y|\tilde{x},\tilde{y}):=f(y^{-1} x y, y^{-1} \tilde{y}) \delta_x(\tilde{x}),
\ee
One can now multiply the kernels $K_f$
in the usual way, and integrate over the gluing variables. Thus, it is easy to check that:
\be
\int dpdq\,\, K_{f_1}(x,y|p,q) K_{f_2}(p,q|\tilde{x},\tilde{y})=K_{f_1\bullet f_2}(x,y|\tilde{x},\tilde{y}).
\ee
However, the kernel formalism leads to a singularity for the Haar
functional $h$. Indeed, we have:
\be
\int dx dy\,\, K_f(x,y|x,y) = \delta_e(e) h(f) = \eta^{-2} h(f).
\ee
Thus, in order to reproduce the correct Haar functional the trace of
the kernel should be renormalized by the already familiar quantity
$\eta^2$ \eqref{eta}.

\subsection{Group field theory Feynman diagram analysis}

Let us first apply the kernel technique to the model that was
constructed from the identity operator $\delta_e(y)$ and that gives
3-manifold invariants. In this subsection we group quantities
according to GFT Feynman graphs $D$. In the next subsection we perform
a ``dual'' analysis, in which a point particle interpretation is more
clear. 

It is clear that the amplitudes can be computed as follows: one should
take the kernel for each vertex of $D$, and compose them together as
specified by the diagram. This is due to the fact that the propagator
of the model is proportional to $\theta$, and the vertex in
\eqref{action*} is invariant under multiplication by $\theta$ from  
any of the 4-possible directions, see \eqref{action*}. Let us analyze
this vertex kernel and give its geometrical
interpretation. 

First we need an expression for the $\theta$-projector
kernel. Applying \eqref{k-2} to \eqref{theta*} we get:
\be\label{k-3}
K_\theta(x_1,y_1;x_2,y_2,x_3,y_3|\tilde{x}_1,\tilde{y}_1;\tilde{x}_2,\tilde{y}_2;\tilde{x}_3,\tilde{y}_3)=
\delta_{x_1}(\tilde{x}_1) \delta_{x_2}(\tilde{x}_2)
\delta_{x_3}(\tilde{x}_3) \\ \nonumber
\delta_e(y_1^{-1} x_1 y_1 y_2^{-1} x_2 y_2 y_3^{-1} x_3 y_3)
\theta(y_1^{-1} \tilde{y}_1,y_2^{-1} \tilde{y}_2,y_3^{-1}
\tilde{y}_3).
\ee
Let us now find the vertex kernel. As in section \ref{sec:Boul}, we
shall enumerate 4 channels of the vertex by indices
$i,j=1,\ldots,4$. The vertex kernel is then a function of variables
$x_{ij}, y_{ij}$, where $x_{ij}\not=x_{ji}, y_{ij}\not=y_{ji}$. The
kernel is obtained by taking a composition of 4 kernels
\eqref{k-3}. We have:
\be
V(x_{ij},y_{ij}) = \int \prod_{i\not=j=1}^4 d\tilde{x}_{ij}
d\tilde{y}_{ij} K_\theta(x_{ij},y_{ij}|\tilde{x}_{ij},\tilde{y}_{ij}),
\ee
where $\tilde{x}_{ij}=\tilde{x}_{ji},
\tilde{y}_{ij}=\tilde{y}_{ji}$. It is straightforward to do the
$\tilde{x}$ integrations. Each kernel \eqref{k-3} contains a $\delta_x(\tilde{x})$
factor, which makes the result proportional to
$\delta_{x_{ij}}(x_{ji})$. Thus, the vertex kernel is actually a
function of 6 variables $x_{ij}=x_{ji}$:
\be\label{k-4}
V(x_{ij},y_{ij}) = \prod_{(i,j,k,l)} \delta_e( y^{-1}_{ij}
x_{ij} y_{ij} y^{-1}_{ik} x_{ik} y_{ik} y^{-1}_{il} x_{il} y_{il}) \\
\nonumber 
\int d\tilde{y}_{ij} \, \theta(y^{-1}_{ij} \tilde{y}_{ij},y^{-1}_{ik}
\tilde{y}_{ik}, y^{-1}_{il} \tilde{y}_{il}).
\ee
Here $\tilde{y}_{ij}=\tilde{y}_{ji}$, so the integration is taken over
6 variables. We have also introduced a notation: $(i,j,k,l)$ for a
quadruple of integers: $i\not=j\not=k\not=l$. The last $\tilde{y}$
integral in \eqref{k-4} can be taken at the
expense of introducing another 4 variables $g_i$. Indeed, let us use
the formula \eqref{theta} for each of the 4 $\theta$-functionals
in \eqref{k-4}. We get:
\be\label{vertex-kern}
V(x_{ij},y_{ij}) = \!\!\!\!\prod_{(i,j,k,l)} \!\!\delta_e( y^{-1}_{ij}
x_{ij} y_{ij} y^{-1}_{ik} x_{ik} y_{ik} y^{-1}_{il} x_{il} y_{il}) 
\int \prod_{i<j} dg_i \,\, \delta_e(y_{ij} g_i g_j^{-1} y^{-1}_{ji}).
\ee
We note that the last term in this expression is just the vertex
kernel of the Boulatov model. Thus, the modification of the vertex as
compared to the Boulatov model case is in an appearance of the 4
additional $\delta$-function terms, and in the fact that the kernel
depends on 6 variables $x_{ij}$. 

One can re-interpret this structure
by saying that the only modification as compared to the Boulatov model
case is that the edge contribution changed: there is now an extra
$\delta$-function for each edge of $D$. Let us change the notation
to display the structure arising more clearly. Let us denote the
vertices, edges and faces of $D$ by tilded letters: $\tilde{v},
\tilde{e}, \tilde{f}$ correspondingly. We leave the untilded letters
for elements of the triangulation $\T$ to which $D$ is dual.
Let us now denote the
$y$-variables as $y_{\tilde{e},\tilde{f}}$, where $\tilde{e}$ is an
edge, and $\tilde{f}$  is a face of
$D$ (each edge $\tilde{e}$ is shared by exactly 3 faces). Let us denote the
$x_{ij}$ variables by $x_{\tilde{f}}$. Indeed, because there is a
$\delta$-function imposing $x_{ij}=x_{ji}$, all $x$-variables
belonging to edges around a face are equal, so there is just one
$x$-variable per face. Thus, the edge factor becomes:
\be\label{edge}
\delta_e(\prod_{\tilde{f}} y^{-1}_{\tilde{e},\tilde{f}} x_{\tilde{f}}^{-1} y_{\tilde{e},\tilde{f}})
\ee
for each edge $\tilde{e}$ of $D$. The vertex, on the other hand, is given by the
same expression as in the case of Boulatov model:
\be\label{vertex-boul} 
\int \prod_{\tilde{e},\tilde{e}'\in \tilde{v}} dg_{\tilde{v},\tilde{e}} \,\,
\delta_e(y_{\tilde{e},\tilde{f}}  g_{\tilde{v},\tilde{e}} 
g_{\tilde{v},\tilde{e}'}^{-1} y^{-1}_{\tilde{e}',\tilde{f}}),
\ee
where for each pair $\tilde{e},\tilde{e}'$ of edges that emanate from vertex $\tilde{v}$ $\tilde{f}$
is the face that shares both $\tilde{e},\tilde{e}'$. Note that the
$g$-variables are different at different vertices
$g_{\tilde{v},\tilde{e}}\not=g_{\tilde{v}',\tilde{e}}$. To obtain the amplitude one 
multiplies the edge and vertex kernels, and integrates over the
$y_{\tilde{e},\tilde{f}}, x_{\tilde{f}}$ variables.  

Already at this stage an interesting geometric interpretation of all
the variables is possible. Indeed, we have already encountered the
$g_{\tilde{v},\tilde{e}}$ variables before. Recall that each Feynman diagram $D$ is dual to
a triangulated 3-manifold, and therefore each vertex $\tilde{v}$ is dual to a
tetrahedron $t$. Thus, all the variables in \eqref{vertex-boul} are those
of a single tetrahedron. Let us denote the triangulation to which $D$
is the dual graph by $\T$. We denote the elements of $\T$ by untilded
letters. Thus, $v,e,f,t$ are vertices,
edges, faces and tetrahedra of $\T$ correspondingly. We see that
$g_{\tilde{v},\tilde{e}}$ are variables of a tetrahedron $t$ dual to
$\tilde{v}$ and correspond to 
faces $f$ of $\T$. Each $g_{\tilde{v},\tilde{e}}$ has the interpretation of
the holonomy  $g^t_f$ of the connection
across the face $f$. There is a similar variable in the neighboring
tetrahedron $t'$ and the total holonomy across the face $f$ is given by the
product: $g_f=(g^t_f)^{-1} g^{t'}_f$.  As we have already discussed in the
introduction, the product $(g^t_f)^{-1} g^t_{f'}$ describes a
rotation of one face $f$ into the other $f'$ and thus carries
information about the dihedral angle between these faces. 

\begin{figure}\label{fig:2tet}
\begin{center}
\epsfig{figure=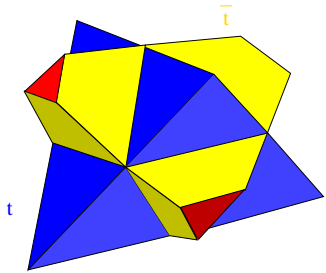, height=1.4in}
\end{center}
\caption{A compound of a tetrahedron $t$ with a truncated tetrahedron $\bar{t}$.}
\end{figure}

The variables $y_{\tilde{e},\tilde{f}}$ are more interesting and this is the first time
that we have encountered them. Using the duality, we can also write
them as $y_{e,f}$. There is 12 of them for each 
tetrahedron. It is natural to interpret
them in terms of a certain other truncated tetrahedron. Indeed, let us introduce
the dual tetrahedron, and cut off its vertices by surfaces 
parallel to the faces of $t$ to obtain a truncated tetrahedron 
$\bar{t}$. A compound object made of $t$ and 
$\bar{t}$ is shown in Fig. \ref{fig:2tet}. The truncated tetrahedron $\bar{t}$ 
has 12 vertices, and we shall interpret 12 variables $y_{e,f}$ as
describing ``positions'' of these vertices. The meaning of each 
$\delta$-function in \eqref{vertex-boul} is then as follows: it imposes the constraint that
the rotation of a vertex $y_{e,f}$ into
$y_{e,f'}$ around the edge $e$ is the same as rotation of the face $f$
into $f'$ as described by $g$-variables.

Thus, we
only have the variables $x_{\tilde{f}}$ uninterpreted. By duality,
they are variables $x_e$ that have to do with the edges of $t$. Tetrahedron
$t$ is truncated, and we interpret $x_e$ as describing rotations of the small
faces of the truncated tetrahedron $t$ into one another. Thus,
interpretation of all of the variables in \eqref{vertex-kern} is in
terms of geometrical quantities associated with two truncated
tetrahedra $t, \bar{t}$ inscribed into one another. 

For the convenience of the reader, let us re-write the above edge and
vertex factors using the triangulation $\T$ notations. We have:
\be\label{edge'}
\delta_e(\prod_e y^{-1}_{e,f} x_e^{-1} y_{e,f})
\ee
for each face $f$ of $\T$. Here the product in the argument of the
$\delta$-function is taken over 3 edges that form the boundary of face
$f$. The tetrahedron factor is given by:
\be\label{vertex-boul'} 
\int \prod_f dg^t_f \prod_{f,f'\in t} \,\,
\delta_e(y_{e,f}  g^t_f 
(g^t_{f'})^{-1} y^{-1}_{e,f'}).
\ee

The interpretation that we would like to propose for the total
tetrahedron amplitude 
\eqref{vertex-kern} is that of a wave function of a pair of dual to
each other truncated tetrahedra. To obtain an amplitude for a manifold
one multiplies these wave functions and integrates over the gluing
variables $x,y$. We would like to note that there is a similarity in
the structure of the vertex found and the expression for the
$6j$-symbol obtained in \cite{Holography}. In both cases the quantity
in question has to do with a truncated tetrahedron, and is constructed
using propagators for long edges (the second set of $\delta$-functions
in \eqref{vertex-kern}), and a set of factors for the small faces (the
first set of $\delta$-functions in \eqref{vertex-kern}). Thus, the
structure we have found is probably pertinent to a very general class
of models.

If not for the edge factors \eqref{edge'}, one could easily integrate over the $y$ variables with the result
being a product of $\delta$-functions, one for every edge $e$ of $\T$, and requiring that the holonomy
around $e$ is trivial. With factors \eqref{edge'} the integration is not straightforward. The total
amplitude \eqref{vertex-kern} is one where tetrahedron constraints discussed in the introduction are taken
into account.

\subsection{Dual graph analysis}

Above we have represented the chain mail evaluation $\Omega CH(D)$ as
a composition of vertex kernels \eqref{vertex-kern}, or the edge
\eqref{edge} and vertex \eqref{vertex-boul} factors. Edges and
vertices are those of $D$. We would now like to group the factors
according not to $D$ but to a triangulation $\T$ whose dual graph is
$D$. 

Thus, Feynman diagrams $D$ have an
interpretation of the dual skeleton of some triangulated
3-manifold. Let us denote this triangulation by $\T$. Let us consider
vertices $v$ of this
triangulation. The edges $e$ of $\T$ are formed by parts of the meridian
links of $CH(D)$. The number of strands forming each edge of $\T$
is not fixed, and is equal to the number of faces of $\T$ sharing
this edge. Strands forming the edges of Feynman diagram $D$, and whose
closure defines the faces of $D$, now enclose the strands forming the
edges of $\T$. Thus, we have a very similar structure to that
considered above, except that it is a dual one, and that the number of
strands enclosed by a link is now not fixed. To evaluate $CH(D)$ we
have to insert the identity operator in each strand. Thus, we insert
in each edge $e$ of $\T$ the
following operator:
\be
\delta_e(x_1^{-1}\ldots x_n^{-1}) \int dx\, \prod_i
\delta_e(x^{-1} y_i).
\ee
As before, it is convenient to introduce the kernel \eqref{k-2}:
\be\label{int-1}
\delta_e(y_1^{-1} x_1^{-1} y_1 \ldots y_n^{-1}
x_n^{-1}y_n) \int dx\, \prod_i
\delta_x(y_i^{-1} \tilde{y}_i) \delta_{x_i}(\tilde{x}_i).
\ee
As in the previous subsection we can take a composition of such
kernels around every vertex of $\T$ by integrating over the
$\tilde{x},\tilde{y}$ variables. Taking into
account the $\delta_{x_i}(\tilde{x}_i)$ functions, we see that there
remains a single variable $x_f$ for every face. The $y$-variables can be
denoted by $y_{e,f}$. Indeed, there is one such variable for each
strand forming each edge of $\T$, or, equivalently,
for each face $f$ sharing edge $e$. This vertex kernel is found to be:
\be\label{v-kern}
\prod_e \delta_e(\prod_{f} y_{e,f}^{-1} x_f^{-1} y_{e,f})
\prod_{e,e'} \delta_e(y_{e,f} x_{v,e} x_{v,e'}^{-1} y_{e',f}^{-1}). 
\ee
Here the second product is over pairs of edges that emanate from $v$,
and $f$ in the argument of the second set of $\delta$-functions is the face
shared by both $e,e'$. Equivalently, one can say that second first
product is over the faces that touch $v$. 
To obtain the chain mail evaluation one should multiply the
kernels \eqref{v-kern} and integrate over all the variables $x_{v,e}, x_f,
y_{e,f}$. Note
that the vertex \eqref{v-kern} is essentially the same as
\eqref{vertex-kern}, except that the dual quantities are used.

Let us discuss a geometrical interpretation of the quantities involved
in \eqref{v-kern}. The quantity $x_f$ is interpreted as the
holonomy across the face $f$. The quantity $y_{e,f}$ encodes the
dihedral angles of a tetrahedron. Namely, $y_{e,f} y_{e,f'}^{-1}$
describes the rotation of face $f$ into $f'$ around edge $e$. We see
that arguments of the first set of $\delta$-functions in \eqref{v-kern} are
just the product of holonomies across the faces that share $e$ with the group
elements that describe rotations of $f$ to the next face $f'$. Thus,
the argument of the $\delta$-function for edge $e$ is the total
rotation that one gets by going around  
$e$. This means that there is no deficit angle allowed! Thus, in spite
of the fact that the model was constructed using $\D$, not just
$\SU(2)$, the model does not
seem to describe point particles. We shall return to this conundrum
below. 

For now let us note that an interpretation of the second set of $\delta$-functions in
\eqref{v-kern} is also possible. Indeed, we see that the quantity $x_{v,e}
x_{v,e'}^{-1}$ describes the rotation of edge $e$ into $e'$, with the
vertex $v$ as the center of rotation. Each $\delta$-function
guarantees that $y_{e,f}^{-1} y_{e',f}$ is the same as the rotation
that sends $e$ to $e'$. 

\subsection{General models and their point particle interpretation}

As we have just seen, the topological GFT model constructed from the
identity operator $\delta_e(y)$ on the Drinfeld double does not
describe point particles. Earlier we have introduced a more general
set of models in which a function $\delta_e(x) P(y)$ is inserted into the
merideans. Now we would like to show that these models do describe
point particles and $P(y)$ is the particle's propagator in the
momentum space. 

We have seen in the previous subsection, the point particle
interpretation, if any, is most visible in a description in which one
groups all the projectors around vertices of $\T$. Let us give this
description. As before, each
edge of $\T$ is formed by a number of strands with the identity
operator inserted. However, the operator on the link that encloses the
edge is now $P$. Therefore, each edge of $\T$ corresponds to the
following operator:
\be
P(x_1^{-1}\ldots x_n^{-1}) \prod_i \delta_e(y_i).
\ee
These operators now have to be multiplied (using the $\bullet$-product) as specified by the fat 
structure, and the Haar functional has to be applied in each strand to get the amplitude. It is
easiest to understand what the result is by noticing that each of these edge operators is 
actually an element of the ideal $({\mathcal A}^*)^{\otimes n}$. Indeed, functions of the form
$f(x) \delta_e(y)$ form an ideal in $D(G)$, isomorphic to the algebra ${\mathcal A}^*$ with
point-wise multiplication. Therefore, the result of the $\bullet$-product of all
these operators is the point-wise product of the functions $P(x_1^{-1}\ldots x_n^{-1})$
times the product of the $\delta$-functions $\delta_e(y)$. It is now clear what the evaluation
is. Let us first discuss what happens with the $y$-dependence. Applying the Haar functional
$h$ in each strand we get an infinity $\delta_e(e)$. These infinities are cancelled
by the prefactor $\eta^{2n_1}$ that we introduced in (\ref{gen-model}). One could,
actually, avoid this infinities altogether by a slight modification of the evaluation procedure.
Thus, since the result in each strand is always in ${\mathcal A}^*$ ideal of $D(G)$, let us
instead of applying the Haar functional of $D(G)$ apply the one of ${\mathcal A}^*$.
The result of this evaluation procedure is finite (for sufficiently nice propagators for which 
the group integrals would converge). 

The result of this evaluation procedure is given by
\be
\int \prod_f dx_f \prod_e P(x_{f_1}^{-1} \ldots x_{f_n}^{-1}).
\ee
Here the first product is over the faces of the diagram, and $x_f$ is a group variable
that gets associated to each face. The second product is over the edges of the diagram,
and the product of $x$-variables in the argument of the propagator includes these variables
for the faces $f_1, \ldots, f_n$ shared by that edge. It is clear now that the function
$P(y)$ plays the role of the particle propagator, for the above answer for the amplitude is as
expected, see e.g. \cite{Barrett,Freidel}.

Thus, we have arrived at the expression we started from in the
Introduction. Indeed, the amplitude is just the Feynman amplitude in
the momentum representation, with the propagator evaluated at
the product of face variables around an edge. The momentum
conservation constraints are solved by introducing the face
variables. The tetrahedron constraints that we discussed in the
Introduction can be omitted because the integrand is invariant under
the action of the gauge transformations at vertices of the dual
triangulation, as can be easily checked. Moreover, as the proposition
\ref{prop*} states, the Feynman amplitude is to a large extent
independent of the extra ``fat'' structure that is added to the
diagram in order to define it. 

There are two interesting ``limiting cases'' to consider. The case $P(y)=\delta_e(y)$
corresponds to having no particles at all. This gives the topological Ponzano-Regge model,
if one introduces an additional prefactor $\eta^{2n_3}$ that is required to guarantee
the invariance under the 2-3 moves. The other case is $P(y)=1$, which gives another
(trivial) topological model, the one that gives one for every manifold as the amplitude.
However, one can decompose the function $P(y)=1$ into $\delta$-functions on
conjugacy classes as in (\ref{ident-1}). This shows that even this trivial model
can be thought of as containing particles of all possible masses in it, but
becoming trivial after the integral over the particles' mass is taken.

\section{Discussion}

Let us recap the main points of this work. We have considered the group
field theory for the Drinfeld double and constructed an exact analog
of the Boulatov model. The Feynman expansion of the model leads to
triangulated 3-manifolds weighted with a manifold invariant. The
invariant is the ``Ponzano-Regge'' model for the Drinfeld
double. By reducing this Feynman amplitude to the chain mail
evaluation we proved that this Drinfeld double Ponzano-Regge amplitude
is indeed an invariant of 3-manifolds (when dressed by an appropriate
power of the formal quantity $\eta^2$). We have also found that this
model does not describe point particles. Indeed, one of the factors
appearing in the amplitude for this model is a $\delta$-function for
the holonomy around each triangulation edge. This is a bit
disappointing, because our original motivation for considering the
Drinfeld double GFT was precisely to obtain a model containing point
particles. Thus, in a certain sense, our original motivation has
failed. 

Nevertheless, motivated by the GFT considerations, and especially by the chain mail
construction, we have introduced a more general set of models that are
defined using not GFT but directly in terms of the chain mail. We have considered only one
such model, but others are possible. The data that goes into
this model is an assignment of a propagator (function on the
$\SU(2)$ group manifold) to each edge of the Feynman diagram. This
propagator can be chosen different for different edges, as long as it
is appropriately normalized. The resulting Feynman amplitude is invariant under moves
that do not change the Feynman graph and the 3-manifold topology. This model
describes point particles. In fact, it is exactly the 
naive model envisaged already in the Introduction. Our results on its
invariance under the topological moves show that this model is
well-defined and to a large extent independent on the ``fat''
structure that was introduced to define it. The model 
models become topological if one choses $P(y)=1$. In this case it seems trivial
as the amplitude it gives to any manifold is one. At the same time,
even in this case the model does describe point particles. One
just has to interpret the operator $\hat{1}$ as the integral over
particle's mass, or using physics terminology, the integral over the
internal momenta of particles. 

The model we have considered (the one constructed with a general
propagator $P$ inserted into the longitudes) has an interpretation as
describing pure 3d gravity coupled to
point particles.  As we have discussed in the introduction, there is
another possible interpretation. Namely, one can view this model
as describing point particles only, but with the backreaction on the
metric taken into account. We have not discussed what
would be the corresponding theory in the position space. Let us only
note that non-commutativity of momentum implies non-commutativity in the position
space as well. It would be of interest to formulate
point particle theories described here in the position space, and understand what
kind of action principle is responsible for them. 

Now that one has a model for 3d quantum gravity with point particles, various
quantities of physical interest can be computed. However, certain
additional developments are necessary in order to achieve this. First it is
important to be able to compute not only closed but also open Feynman graph
amplitudes. This would give correlation
functions. Second and more important, one should find an analog of LSZ
reduction formula of field theory that expresses scattering amplitudes
in terms of field correlation functions. We shall not attempt this in the present paper. 

A remark is in order about the topological invariant defined by our particle model. 
As we have already mentioned, this is ${\mathcal Z}=1$ for any manifold. However, the
amplitude for each Feynman diagram (with a fixed set of conjugacy classes
$\theta_e$ on the edges) is non trivial.  Moreover, an amplitude for
an open Feynman graph, which describes particle scattering and which
we have not considered in this paper is also not trivial, even when
one integrates over the internal momenta. This shows that, in spite of
seeming triviality, this model is physically interesting.

Let us now return to the topologically invariant model constructed from the
identity operator $\delta_e(y)$. Several comments are in order. First,
it would be interesting to understand
how different this model is from that of Ponzano-Regge. 
The original Ponzano-Regge model imposes
the constraints that the holonomy either around edges or dual edges is trivial. The model constructed
in this paper imposes both sets of constraints at the same time. For
the case of the double of a quantum $\SU_q(2)$, which makes the model
well-defined (by rendering $\eta^{-2}$ finite), one expects the
Drinfeld double ``Ponzano-Regge'' model to be equal to the usual
Ponzano-Regge model amplitude squared. It would be interesting to
understand the interpretation of the classical group Drinfeld double
case better. 

Another remark is that, as is clear from our construction, the gauge symmetry
$\delta\phi=\psi, \theta\bullet\psi=0$ of the group
field theory is extremely important. From the chain mail description
it is clear  that topological invariance of the model is intimately
related to the fact that $\theta$-projector
is used as the propagator of the model. On the other hand, we know
that topological invariance of 3d gravity is due to large amount of
gauge symmetry present in it. More precisely, diffeomorphisms are so
strong in 3d, that they render the theory topological. All this
strongly suggests that there is a link between the group field theory
and gravity gauge symmetries. In the context of 2d gravity matrix
models the relation between gravity and a matrix model is an instance
of open-closed string duality, in which one and the same Feynman
amplitude has two different interpretations. We expect that the
relation between group field theory and 3d gravity explored here is of
a similar nature, and that the group field theory gauge symmetry
becomes diffeomorphism symmetry on the gravity side. It would be of
considerable interest to explore all this in more detail. 

It is clear that the GFT construction that we have presented is very
general, and can be applied to algebras other than the Drinfeld
double. The Drinfeld double of $\SU(2)$ is described \cite{Koorn} as the twisted product of the
algebra of functions on $\SU(2)$ with its dual, which is also algebra
of functions on $\SU(2)$. The Drinfeld twisted product construction
can be applied to groups other than $\SU(2)$. Another important for
physical applications example is given by the quantum Lorentz group,
which is just the twisted product of the quantum $\SU(2)$ with $q$
real and its Pontryagin dual, see \cite{Woronowicz}. Our construction
can be applied to this case. We expect that in this case there is not
one, but several possible models that lead to 3-manifold invariants,
depending on which ``identity'' operator is used. These models should
have the physical interpretation as describing point particles coupled
to quantum gravity with either positive cosmological constant and
Lorentzian signature or with negative cosmological constant and
Euclidean signature. It would be of great interest to find all these
models.

\section*{Acknowledgments}

I would like to thank my colleagues John Barrett and Jorma Louko for
stimulating discussions and support. I am grateful to G. 't Hooft
for a discussion about point particles in 2+1 gravity, and to
L. Freidel for many important discussions about group field theory.
The author is supported by an EPSRC Advanced Fellowship.

\section*{Appendix: Some useful formulas}

In the main text we often use the integral of two and three matrix elements. For two matrix elements we have:
\be\label{A-2-1}
\int dg\,\, \overline{T^j_{mn}(g)} T^{j'}_{m'n'}(g) = \frac{\delta_{jj'}}{{\rm dim}_j} \delta_{mm'} \delta_{nn'}.
\ee
In graphical notation
this takes the following form:
\be\label{A-2}
\int dg \,\,\, \lower0.25in\hbox{\epsfig{figure=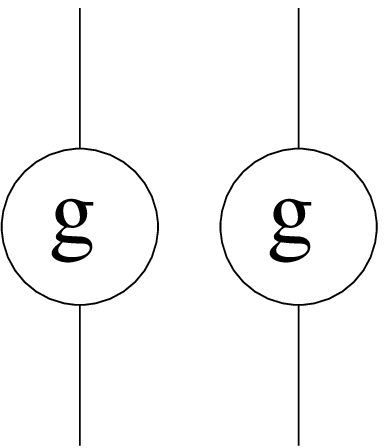, height=0.5in}} = \frac{1}{{\rm dim}}\,\,
\lower0.25in\hbox{\epsfig{figure=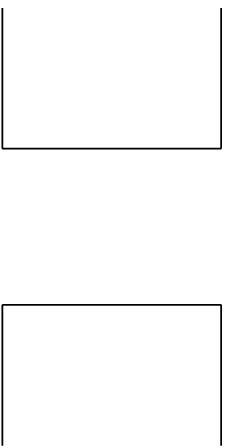, height=0.5in}}
\ee
and
\be\label{A-3}
\int dg \,\,\, \lower0.25in\hbox{\epsfig{figure=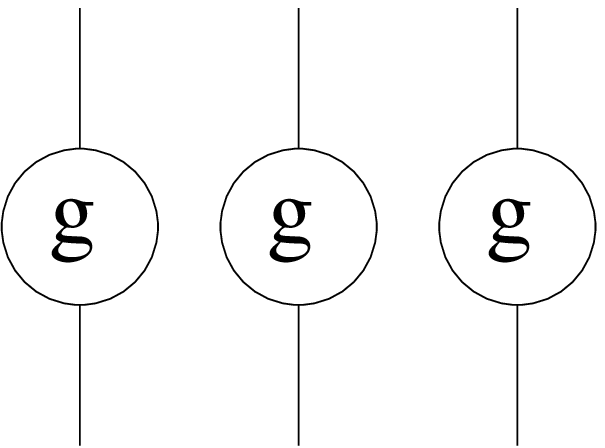, height=0.5in}} =
\lower0.25in\hbox{\epsfig{figure=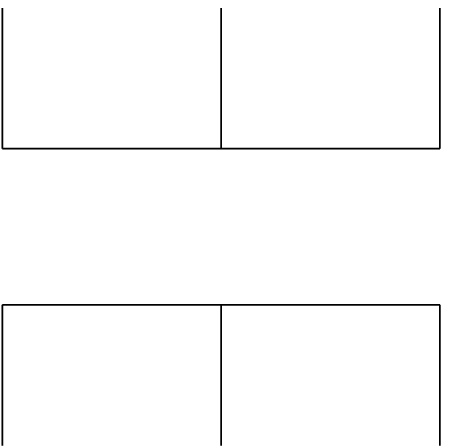, height=0.5in}}
\ee
The intertwiner that is used in the last formula is chosen to be normalized so that:
\be\label{A-theta}
\lower0.35in\hbox{\epsfig{figure=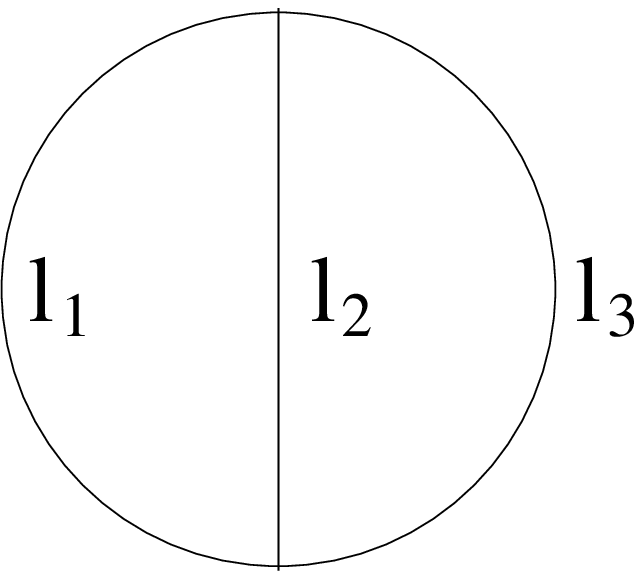, height=0.7in}} = 1
\ee
whenever the 3 spins satisfy the triangular inequalities, and zero otherwise.


\begin{thebibliography}{99}

\bibitem{Barrett}
  J.~W.~Barrett,
  ``Feynman diagrams coupled to three-dimensional quantum gravity,''
  arXiv:gr-qc/0502048.


\bibitem{Freidel}
  L.~Freidel and E.~R.~Livine,
  ``Ponzano-Regge model revisited. III: Feynman diagrams and effective field
  theory,''
  arXiv:hep-th/0502106.


\bibitem{PR} G. Ponzano and T. Regge, ``Semiclassical limit of Racah coefficients'' 
in Spectroscopic and group theoretical methods in physics (Bloch ed.),
North-Holland, 1968.

\bibitem{Freidel-div}
  L.~Freidel and D.~Louapre,
  ``Diffeomorphisms and spin foam models,''
  Nucl.\ Phys.\ B {\bf 662}, 279 (2003)
  [arXiv:gr-qc/0212001].

\bibitem{Majid-F}
  L.~Freidel and S.~Majid,
  ``Noncommutative harmonic analysis, sampling theory and the Duflo map in 2+1
  quantum gravity,''
  arXiv:hep-th/0601004.

\bibitem{Freidel-CS}
  L.~Freidel and D.~Louapre,
  ``Ponzano-Regge model revisited. II: Equivalence with Chern-Simons,''
  arXiv:gr-qc/0410141.

\bibitem{Barrett-Obs}
  J.~W.~Barrett, J.~M.~Garcia-Islas and J.~F.~Martins,
  ``Observables in the Turaev-Viro and Crane-Yetter models,''
  arXiv:math.qa/0411281.

\bibitem{Boulatov}
  D.~V.~Boulatov,
  ``A Model of three-dimensional lattice gravity,''
  Mod.\ Phys.\ Lett.\ A {\bf 7}, 1629 (1992)
  [arXiv:hep-th/9202074].


\bibitem{MM}
  P.~H.~Ginsparg and G.~W.~Moore,
  ``Lectures on 2-D gravity and 2-D string theory,''
  arXiv:hep-th/9304011.



\bibitem{Sum}
  L.~Freidel and D.~Louapre,
  ``Non-perturbative summation over 3D discrete topologies,''
  Phys.\ Rev.\ D {\bf 68}, 104004 (2003)
  [arXiv:hep-th/0211026].

\bibitem{BM}
  F.~A.~Bais and N.~M.~Muller,
  ``Topological field theory and the quantum double of SU(2),''
  Nucl.\ Phys.\ B {\bf 530}, 349 (1998)
  [arXiv:hep-th/9804130].




\bibitem{Mikovic}
  A.~Mikovic,
  ``Spin foam models of matter coupled to gravity,''
  Class.\ Quant.\ Grav.\  {\bf 19}, 2335 (2002)
  [arXiv:hep-th/0108099].


\bibitem{Laurent-recent}
  L.~Freidel,
  ``Group Field Theory: An overview,''
  arXiv:hep-th/0505016.

\bibitem{Rovelli}
  M.~P.~Reisenberger and C.~Rovelli,
  ``Spacetime as a Feynman diagram: The connection formulation,''
  Class.\ Quant.\ Grav.\  {\bf 18}, 121 (2001)
  [arXiv:gr-qc/0002095].


\bibitem{Philippe}  R.\ Dijkgraaf, V.\ Pasquier and Ph.\ Roche, 
``Quasi-Hopf Algebras, Group Cohomology and Orbifold models,''
  Nucl.\ Phys.\ Proc Suppl.\ {\bf 18B},
 60-72 (1990)




\bibitem{Koorn}
  T.~H.~Koornwinder, B.~J.~Schroers, J.~K.~Slingerland and F.~A.~Bais,
  ``Fourier transform and the Verlinde formula for the quantum double of a
  finite group,''
  J.\ Phys.\ A {\bf 32}, 8539 (1999)
  [arXiv:math.qa/9904029].


\bibitem{Roberts} J.~Roberts, ``Skein theory and Turaev-Viro invariants'',
Topology {\bf 34} 771-787 (1995).

\bibitem{Roberto} 
  R.~De Pietri and C.~Petronio,
  ``Feynman diagrams of generalized matrix models and the associated  manifolds
  in dimension 4,''
  J.\ Math.\ Phys.\  {\bf 41}, 6671 (2000)
  [arXiv:gr-qc/0004045].

\bibitem{Woronowicz} S. L. Woronowicz and S. Zakrzewski, ``Quantum
  deformations of the Lorentz group. The *-algebra level'', Compositio
  Math.\ {\bf 90} 211-243 (1994).

\bibitem{Holography} K. Krasnov, ``Holography for the Lorentz group
  Racah coefficients'', Class.\ Quant.\ Grav.\ {\bf 22} 1933-1944 (2005).

\end{thebibliography}
\end{document}